\documentclass[11pt,a4paper]{article}
\usepackage{amsmath,graphicx,amssymb}
 \usepackage[colorlinks=true]{hyperref}
\hypersetup{urlcolor=blue, citecolor=red}

\usepackage{caption,subcaption,cite}

\newcommand\figwidth{.82\textwidth }

\usepackage[utf8]{inputenc}
\newcommand\tsr\mathbb
\newcommand\ve{\vec e}
\newcommand\e{\textrm{e}}
\newcommand\dtimes{\dot{\times}}
\newcommand\di\displaystyle
\newcommand{\matriced}[4]{\begin{pmatrix} #1 & #2 \\ #3 & #4\end{pmatrix}}
\newcommand{\vecteurd}[2]{\begin{pmatrix} #1 \\ #2 \end{pmatrix}}
\DeclareMathOperator{\grad}{grad}
\DeclareMathOperator{\dive}{div}
\DeclareMathOperator{\rot}{curl}

\makeatletter
\@namedef{u8:\detokenize{β}}{\ensuremath{\beta}}
\@namedef{u8:\detokenize{γ}}{\ensuremath{\gamma}}
\@namedef{u8:\detokenize{Δ}}{\ensuremath{\Delta}}
\@namedef{u8:\detokenize{θ}}{\ensuremath{\theta}}
\@namedef{u8:\detokenize{κ}}{\ensuremath{\kappa}}
\@namedef{u8:\detokenize{λ}}{\ensuremath{\lambda}}
\@namedef{u8:\detokenize{μ}}{\ensuremath{\mu}} % N'enlever aucune de ces 2 lignes
\@namedef{u8:\detokenize{µ}}{\ensuremath{\mu}} % N'enlever aucune de ces 2 lignes
\@namedef{u8:\detokenize{ν}}{\ensuremath{\nu}}
\@namedef{u8:\detokenize{ξ}}{\ensuremath{\xi}}
\@namedef{u8:\detokenize{π}}{\ensuremath{\pi}}
\@namedef{u8:\detokenize{ρ}}{\ensuremath{\rho}}
\@namedef{u8:\detokenize{σ}}{\ensuremath{\sigma}}
\@namedef{u8:\detokenize{τ}}{\ensuremath{\tau}}
\@namedef{u8:\detokenize{ψ}}{\ensuremath{\psi}}
\@namedef{u8:\detokenize{Ω}}{\ensuremath{\Omega}}
\@namedef{u8:\detokenize{ω}}{\ensuremath{\omega}}
\@namedef{u8:\detokenize{÷}}{\frac}
\@namedef{u8:\detokenize{∂}}{\ensuremath{\partial}}
\@namedef{u8:\detokenize{ｄ}}{\ensuremath{{\,\mathrm{d}}}}
\@namedef{u8:\detokenize{ℤ}}{\ensuremath{\mathbb{Z}}}
\@namedef{u8:\detokenize{⋀}}{\ensuremath{\wedge}}
\@namedef{u8:\detokenize{∫}}{\ensuremath{\int}}
\@namedef{u8:\detokenize{∑}}{\ensuremath{\sum}}
\@namedef{u8:\detokenize{ℝ}}{\ensuremath{\mathbb{R}}}
\@namedef{u8:\detokenize{ℕ}}{\ensuremath{\mathbb{N}}}
\@namedef{u8:\detokenize{≥}}{\ensuremath{\ge}}
\@namedef{u8:\detokenize{⟨}}{\ensuremath{\langle}}
\@namedef{u8:\detokenize{⟩}}{\ensuremath{\rangle}}
\@namedef{u8:\detokenize{‖}}{\ensuremath{\Vert}}
\@namedef{u8:\detokenize{∈}}{\ensuremath{\in}}
\@namedef{u8:\detokenize{√}}{\ensuremath{\sqrt}}
\@namedef{u8:\detokenize{∇}}{\ensuremath{\nabla}}
\@namedef{u8:\detokenize{·}}{\ensuremath{\cdot}}
\@namedef{u8:\detokenize{≠}}{\ensuremath{\ne}}
\@namedef{u8:\detokenize{⦅}}{\left(}
\@namedef{u8:\detokenize{⦆}}{\right)}
\@namedef{u8:\detokenize{⟦}}{\left[}
\@namedef{u8:\detokenize{⟧}}{\right]}
\@namedef{u8:\detokenize{⟪}}{\left\langle}
\@namedef{u8:\detokenize{⟫}}{\right\rangle}
\makeatother

\title{A new Hodge operator in Discrete Exterior Calculus. Application to fluid mechanics}

\author{Rama Ayoub, Aziz Hamdouni, Dina Razafindralandy}
\date{}

\begin{document}
\maketitle
\vspace{-2\baselineskip}
\begin{center}
\small Laboratoire des Sciences de l'Ingénieur pour l'Environnement\\ UMR 7356 La Rochelle Université -- CNRS
\end{center}
\begin{abstract}
	This article introduces a new and general construction of discrete Hodge operator in the context of Discrete Exterior Calculus (DEC). This discrete Hodge operator permits to circumvent the well-centeredness limitation on the mesh with the popular diagonal Hodge. It allows a dual mesh based on any interior point, such as the incenter or the barycenter. It opens the way towards mesh-optimized discrete Hodge operators. In the particular case of a well-centered triangulation, it reduces to the diagonal Hodge if the dual mesh is circumcentric. Based on an analytical development, this discrete Hodge does not make use of Whitney forms, and is exact on piecewise constant forms, whichever interior point is chosen for the construction of the dual mesh.
	Numerical tests oriented to the resolution of fluid mechanics problems and thermal transfer are carried out. Convergence on various types of mesh is investigated. Flat and non-flat domains are considered.
\end{abstract}

\section{Introduction}

The interest for (geometric) structure-preserving numerical integrators has grown in computing community \cite{hairer06,blanes16,amses18,bochev06,arnold10,amses19,salnikov18}. This is due to the ability of these schemes to reproduce fundamental physical properties (conservation laws, \dots) of the equations and their robustness in long-time integration. Discrete Exterior Calculus (DEC) belongs to this family of integrators. It aims at developing a discrete version of the theory of exterior calculus, and more generally the differential geometry framework, where most equations of physics are formulated. Initially  developed by Bossavit for electromagnetism in a series of papers \cite{bossavit98_1,bossavit98_2,bossavit98_3,bossavit98_4,bossavit99_1,bossavit99_2,bossavit99_3,bossavit99_4,bossavit99_5}, DEC was used in fluid mechanics for the resolution of Darcy's equation and the simulation of the dynamics of ideal and viscous fluid flows in basic configurations \cite{elcott07,hirani15,mohamed16a}.

The primary calculus tools of the DEC are discrete differential forms or cochains. An important advantage of DEC is that the Stokes' theorem is naturally verified at discrete level. This is due to the construction of the discrete exterior derivative operator $ｄ$ by duality with the boundary operator. A second advantage is that the DEC framework offers a  naturally coherent discretization 
of derivative operators (divergence, gradient, curl in 2D and 3D) such that the usual relations
\[\rot \grad=0,\quad\quad\text{and}\quad\quad\dive\rot=0\] 
are verified at machine precision. This is due to the fact that these derivative operators are all represented by the exterior derivative $ｄ$ in exterior calculus framework and that in DEC, the discrete $ｄ$ obeys the relation
\begin{equation}
	ｄ^2=0, \label{d2}
\end{equation}
just like the continuous $ｄ$.
These properties of DEC permitted for example Elcott et al. \cite{elcott07} to design a circulation-preserving numerical scheme for the simulation of ideal fluid flows. These properties also avoid the appearance of spurious quantities such as an artificial mass, potential or portance in the numerical solution.

When used as discretization method, DEC can be seen as a finite volume method in exterior calculus. Indeed, in this approach, cochains are differential forms integrated over the different elements of the mesh (vertices, edges, triangles, \dots). Disrete exterior calculus has a finite element version, developed mainly by Arnold and his collaborators in \cite{arnold06,arnold10,arnold18}.

A key operator in exterior calculus, needed to express constitutive laws for example, is the Hodge star operator. In a finite element approach, the discrete Hodge operator is built straightforwardly, by applying the continuous Hodge to the polynomial differential forms which constitute the basis. After a suitable inner product, this results in the mass matrix. In DEC, to which the present article is limited, defining a discrete Hodge is less natural because one has to deal with already intregrated forms.  In particular, in order to ensure the bijectivity of the discrete Hodge, a dual mesh is necessary in DEC, whereas such a notion is not required in the finite element approach. The circumcentric dual is a popular choice of dual mesh, due to the orthogonal nature of simplicial cells and their circumcentric duals. This orthogonality permits a simple construction of a discrete Hodge, having a diagonal matrix representation \cite{bossavit99_2,hirani03}. The diagonal entries of this discrete Hodge, sometimes called circumcentric or diagonal Hodge, are simply the ratios of the volumes of dual cells and primal simplices.

The diagonal Hodge was initialy designed for completely well-centered simplicial meshes, that are meshes where each simplex contains its circumcenter.
In practice, it is difficult to generate this type of triangulation \cite{Rajan94}. It can be done for some specific simple geometries in $\mathbb{R}^2$ \cite{cassidi81,yuan10,zamfirescu13}. For more complex geometries, VanderZee et al. propose a reprocessing algorithm making a 2D mesh well-centered but it is not very practical and may not work in 3D \cite{vanderzee10}. Hirani and his collaborators extended the use of the diagonal Hodge to Delaunay and to pairwise non-Delaunay meshes, by introducing signed elementary dual volumes \cite{Hirani13,mohamed18}. However, this approach may still lead to zero-volume dual edges and cells when positive contributions compensate negative ones. A reprocessing is then necessary. Mullen et al. \cite{Mullen11hot:hodge-optimized} proposed an optimization of the triangulation to the diagonal Hodge. To this aim, they introduce weighted duals and weighted circumcenter, and construct the mesh by minimising the error of the diagonal Hodge. This technique generates a well-shaped mesh, but not well centered. Moreover, it may become expensive, especially for problems with time-varying domains such as fluid-structure interaction. Note also that all these discrete Hodge operators require that the dual mesh is based on the circumcenter. 

Another existing discrete Hodge operator, studied by Bossavit from the beginning of DEC, is the Galerkin Hodge \cite{bossavit98,bossavit99_5}. It corresponds to a mass matrix, with the Whitney forms as shape functions \cite{whitney57,bossavit88}. It is built within a finite element framework. A drawback of using the Galerkin Hodge in (the finite-volume flavoured) DEC is that, not only it introduces an inconsistency but also it is not exact even for constant forms. This can be shown straighforwardly in a right but not isosceles triangle. A more DEC- than finite element-flavoured discrete Hodge, but still based on Whitney forms, is proposed in \cite{tarhasaari99}. It is defined through an interpolation with Whitney forms and an integration on dual simplices and is generally called Whitney Hodge. Another discrete Hodge operator which also uses Whitney forms is proposed in \cite{Auchmann06}. Called geometric Hodge, it does not need an interpolation nor integration concept. For symmetry reasons, it requires a dual mesh based on the barycenters. In fact, as noticed in \cite{mohamed16b} in 2D, the geometric Hodge is a symmetric approximation of the Galerkin Hodge where the integration is approximated by a one point quadrature with evaluation only at the barycenter. Like the usual Galerkin and the Whitney Hodge, the geometric Hodge is represented by a sparse but not diagonal matrix. However, numerical comparisons in \cite{mohamed16b} show that the additional computational time induced by the lack of diagonal structure is not very high.

In the present article, we propose a new construction of a discrete Hodge operator which comply the following requirements. First, it does not need a reprocessing of the primal mesh. This condition is important since, in some situations, modifying the mesh may be harmful. It may for instance affect the shape of surfaces. Moreover, as mentioned, the induced cost increase may become important in time-varying domains. Second, the construction must be general enough such that the centers on which the dual mesh is based can be changed at convenience, as long as the dual mesh has no degenerate element. For example, the construction of the discrete Hodge must be valid for both a barycentric and an incentric duals. Neither the diagonal nor the geometric Hodge is appropriate for this task since with these operators, the dual mesh is imposed. Lastly, the Hodge must be exact on piecewise constant differential forms, whichever choice is made on the dual mesh. 

By proposing such a construction, we intend to develop a discrete Hodge which adapts to the mesh in order to minimize some quantity (the global error, \dots) or to comply to other physical requirements, by choosing algorithmically the most suitable center. These center may even vary from a simplex to another. This optimization process, which will be tackled in a future work, would contrast with that in \cite{Mullen11hot:hodge-optimized} where it is the mesh which adapts to the (diagonal) Hodge operator. 

In section 2, a brief reminder on DEC, on the continuous Hodge operator and on the diagonal Hodge is done. A deeper introduction to DEC can be found in the papers of Bossavit \cite{bossavit98_1,bossavit98_2,bossavit98_3,bossavit98_4,bossavit99_1,bossavit99_2,bossavit99_3,bossavit99_4,bossavit99_5} and in \cite{hirani03,desbrun05,crane13} for instance. The development of the new discrete Hodge, followed by illustrative examples, a basic preliminary error analysis and a discussion, is presented in section 3. We end up with numerical experiments, carried out on Poisson equations and on isothermal and non-isothermal fluid dynamics problems, in section 4. Various types of mesh (well-centered mesh, right triangulation and non-Delaunay meshes) will be considered. Convergence analyses will be carried out. Although our goal is not to compare the new discrete Hodge to existing ones, results given by the diagonal Hodge are presented as reference, when available. An application on a non-flat domain is presented.

\section{Review on DEC}

DEC is a theory of exterior calculus on a discretized domain, which preserves Stokes' theorem and the exactness of relation (\ref{d2}). In this section, we make a reminder on oriented simplicial discretization of a domain. Next, since the primary tools of exterior calculus are differential forms, we briefly present how to discretize them into cochains. Finally, the Hodge star operator and its discretization into the diagonal Hodge is recalled.

The basic notions on differential geometry can be found in many books in literature. One can cite \cite{abraham88,lee12,fecko06,morita01} for a formal introduction and books \cite{spivak991,spivak992,spivak993,spivak994,spivak995} for a more illustrative one. Specifically on DEC, Bossavit's articles \cite{bossavit98_1,bossavit98_2,bossavit98_3,bossavit98_4,bossavit99_1,bossavit99_2,bossavit99_3,bossavit99_4,bossavit99_5} and the thesis \cite{hirani03} can be suggested. The reader can also refered to \cite{crane13} which proposes a lecture on DEC from scratch, from a computation-oriented introduction to differential geometry to program writing. The lectures \cite{grinspun05} on digital differential geometry can also be mentioned.

\subsection{Discretization of the domain\label{sec:discretization_domain}}

\begin{figure}[ht]
	\centering
	\includegraphics[width=4cm]{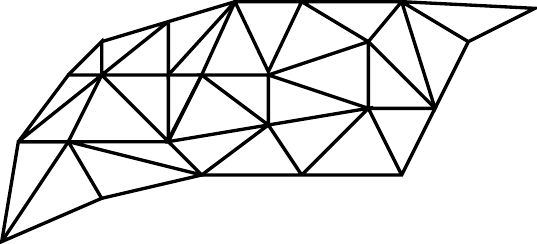}
	\caption{Example of 2D simplicial complex embedded in $ℝ^3$}
	\label{fig:complex}
\end{figure}

Consider a system of partial differential equations defined on an $n$-dimensional spatial domain $M\subset ℝ^m$ for some $m∈ℕ$. The domain $M$ is subdivised into $n$-dimensional simplices. Recall that a $k$-simplex is the convex hull of $k+1$ points. For instance, if $\dim M=3$ then $M$ is subdivised into tetrahedra. We denote $K$ the set composed of these $n$-dimensional simplices, together with their faces, the faces of the faces, and so on untill the vertices. For example, in 3D, $K$ is composed of tetrahedra (3-simplices), triangles (2-simplices), edges (1-simplices) and vertices (0-simplices). In 2D meshes, the top-dimensional simplices are triangles. As an example, Figure \ref{fig:complex} presents a simplicial discretization of a curved surface in $ℝ^3$. 
Some regularity conditions are also generally usefull \cite{hirani03,desbrun05,gillette09}.

\begin{figure}[ht]\centering
	\includegraphics[width=4cm]{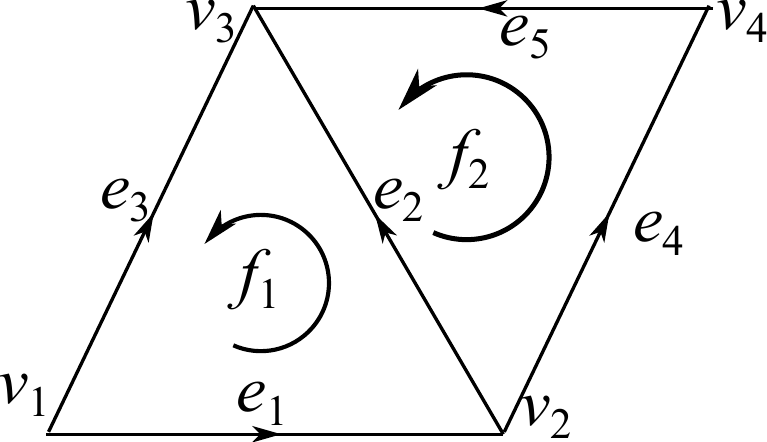}
	\caption{Example of a consistently oriented mesh. Arrows represent the orientation of edges and faces}
	\label{mesh}
\end{figure}

Each element of $K$ is assigned an orientation, which can simply be defined from an ordering of its nodes. Each simplex has two possible orientations. Two adjacent top-dimensional simplices are required to have the same orientation. By contrast, the orientation of each lower dimensional simplex is arbitrary. A sample 2D oriented complex is presented in Figure \ref{mesh}. For simplicity, and since it is sufficient for our applications, we assume that all the vertices have the same (positive) orientation. 

We denote $K_k$ the set of oriented $k$-dimensional simplices of $K$
\begin{equation*}K_k=\{σ\in K,\ σ\text{ oriented},\ \dim σ=k\}\end{equation*}
and $Ω_k$ the vector space spanned by formal linear combination of elements of $K_k$
\begin{equation*}Ω_k=\operatorname{span}K_k=\left\{c=\sum_{σ_i\in K_k}c_iσ_i,\ c_i\in ℝ\right\}.\end{equation*}
An element $c$ of $Ω_k$ is called a $k-$chain. In theory, the components $c_i$ may take any real value but, in practice, only values in $\{0,1,-1\}$ have physical meaning. A value 0 means that the element $σ_i$ does not belong to the chain, 1 means that it is present in $c$, and $-1$ when it is present but with the opposite orientation. As an example, the chain 
\begin{equation}
	e_1+e_4+e_5-e_3\in Ω_1
	\label{chain}
\end{equation}
in Figure \ref{mesh} constitutes a closed loop.

\subsection{Boundary operator}

The boundary of a $k$-dimensional simplex is the sum of its oriented $(k-1)$-dimensional faces. In this sum, each face is given a sign, depending on wether its orientation is consistent with that of the considered $k$-simplex. For example, the boundary of the face $f_1$ in Figure \ref{mesh} is
\begin{equation*}
	∂f_1=e_1+e_2-e_3.
\end{equation*}
The boundary of $e_1$ is
\begin{equation*}
	∂e_1=v_2-v_1.
\end{equation*}
The boundary operator $∂$ extends into a linear map from $Ω_k$ to $Ω_{k-1}$ by defining the boundary of a chain $∑_{i∈I}c_iσ_i$, for some set $I$ of indices, as follows:
\begin{equation}
	∂⟦∑_{i∈I}c_iσ_i⟧=∑_{i∈I}c_i∂σ_i.
\end{equation}%
The operator $∂$ can be represented by a sparse matrix. For instance, the non-zero boundary operators on the mesh in Figure \ref{mesh} are
\begin{equation}
	∂_{|Ω_2}=\begin{pmatrix}
		1&0\\1&-1\\-1&0\\0&1\\0&1
	\end{pmatrix},\quad
	∂_{|Ω_1}=\begin{pmatrix}
		-1& 0&-1& 0& 0&\\
		 1&-1& 0&-1& 0&\\
		 0& 1& 1& 0& 1&\\
		 0& 0& 0& 1&-1&
	\end{pmatrix}.
	\label{boundary}
\end{equation}
In equation (\ref{boundary}), $∂_{|Ω_k}$ is the restriction of $∂$ to $Ω_k$, which acts on $k$-simplices.

\subsection{Discrete differential forms and exterior derivative}

A discrete $k$-form, or a $k-$cochain, is an element of the algebraic dual $Ω^k(K)$ of $Ω_k$. Since the elements of $K_k$ form a basis of $Ω_k$, a discrete $k$-form is simply a map which, to each element of $K_k$, assigns a real number:
\begin{equation}
	ω\in Ω^k(K)\ :\ \begin{array}{ccc}K_k&\longrightarrow& ℝ\\σ&\longmapsto&⟨ω,σ⟩.\end{array}
	\label{primal_cochain}
\end{equation}
Computationally, a $k$-cochain is represented by an array $(⟨ω,σ_i⟩)_{σ_i∈K_k}$.

In DEC, a differential $k$-form $ω\inΩ^k(M)$ on $M$ is discretized into a $k$-cochain $ω_K∈Ω^k(K)$ by taking the pairing $⟨·,·⟩$ as an integration
\[ ⟨ω_K,c⟩=∫_cω,\quad\quad\text{for all }c∈Ω_k\]
if $k≥1$, and as a simple evaluation
\[ ⟨ω_K,c⟩=ω(c)\] 
if $c∈Ω_0$ is a point. 
In the sequel, the cochain $ω_K$ will also be denoted $ω$ when there is no confusion.

The (continuous) exterior derivative ｄ can be locally defined by 
\[ ｄ⦅ω_{i_1\ldots i_k}ｄx^{i_1}⋀\ldots⋀ｄx^{i_k}⦆=\cfrac{∂ω_{i_1\ldots i_k}}{∂x^j}ｄx^j⋀ｄx^{i_1}⋀\ldots⋀ｄx^{i_k}\]
(with a summation on $j$). It verifies Stokes'theorem
\begin{equation}
	∫_Cｄω=∫_{∂C}ω
	\label{eq:stokes}
\end{equation}
for any orientable manifold $C$ with boundary $∂C$ and any $(\dim C)$-form ω. Mimicking relation (\ref{eq:stokes}),
The discrete exterior derivative operator, also denoted $ｄ$, is defined as the dual of the boundary operator $∂$:
\begin{equation}
	⟨ｄω,c⟩=⟨ω,∂c⟩,\quad \text{ for all } ω\inΩ^k(K), c\in K_{k+1}.
	\label{d}
\end{equation}
It is a linear map from $Ω^k$ to $Ω^{k+1}$. %It can be remarked that relation (\ref{d}) would simply be the expression of Stokes' theorem if the pairing $⟨·,·⟩$ was an integration and ω a differential form. 
Computationally, the matrix $ｄ_{|Ω^k(K)}$ of the restriction of $ｄ$ on the space of $k$-cochains is the transpose of the matrix of $∂_{|Ω_{k+1}}$:
\[ｄ_{|Ω^k(K)}=∂_{|Ω_{k+1}}^\mathsf{T}.\]

Since $∂^2=0$ (the boundary of a boundary is the empty set), it follows by the duality relation (\ref{d}) that
\begin{equation}
	ｄ^2=0.
\end{equation}
The following sequence is then, as in the continuous case, exact:
\begin{equation}
	0\xrightarrow{\makebox[5mm]{$ｄ$}}Ω^0(K)\xrightarrow{\makebox[5mm]{$ｄ$}}Ω^1(K)\xrightarrow{\makebox[5mm]{$ｄ$}}\cdots\xrightarrow{\makebox[5mm]{$ｄ$}}Ω^{n-1}(K)\xrightarrow{\makebox[5mm]{$ｄ$}}Ω^n(K)\xrightarrow{\makebox[5mm]{$ｄ$}}0.
	\label{derham}
\end{equation}

\subsection{The diagonal Hodge}

Having applications in the field of fluid mechanics domain in mind, we assume that $M$ is a flat domain in $ℝ^n$, $n=2$ or 3, from now on, to simplify. It inherits the usual Euclidean metric and the induced right-handed oriented volume form of $ℝ^n$. %We also choose a righ-handed orientation.

To express material laws in exterior calculus framework, the Hodge star operator $\star$ is necessary. It realizes an isomorphism between $\Omega^k(M)$ and $\Omega^{n-k}(M)$. Rather than giving its definition (which can be found for example in \cite{marsden99}), we simply recall that, if $n=2$,
\begin{equation}
	\begin{array}{lll}
		\star 1=ｄx\wedge ｄy, \\[7pt]
	\star ｄx=ｄy,\quad&\star ｄy=-ｄx, \\[7pt]
	\star (ｄx\wedge ｄy)=1. 
	\end{array}
	\label{star2d}
\end{equation}
In fact, these relations completely characterize the operator $\star$.
For $n=3$, we have
\begin{equation*}
	\begin{array}{lll}
		\star 1=ｄx\wedge ｄy \wedge ｄz,\\[7pt]
		\star ｄx=ｄy\wedge ｄz,\quad&\star ｄy=ｄz\wedge ｄx,\quad&\star ｄz=ｄx \wedge ｄy,\\[7pt]
		\star(ｄy\wedge ｄz)=ｄx,\quad&\star(ｄz\wedge ｄx)=ｄy,\quad&\star(ｄx \wedge ｄy)=ｄz,\\[7pt]
		\star (ｄx\wedge ｄy\wedge ｄz)=1.
	\end{array}
\end{equation*}
It can be checked that
\begin{equation}
	\star_1\star_1 = -\mathsf{Id}
	\label{starstar}
\end{equation}
in 2D and 
\[ \star_2\star_1 = \mathsf{Id} \]
in 3D, where $\mathsf{Id}$ is the identity map and $\star_k$ is the restriction of $\star$ to $Ω^k(M)$. Note also that the Hodge star verifies the relation
\begin{equation}
	ω(u_1,\dots,u_k)=\star ω(u_{k+1},\dots,u_{n}) \label{ortho}
\end{equation}
for any positively oriented orthonormal basis $(u_1,\dots,u_n)$ of the tangent space and any differential $k$-form $ω∈Ω^k(M)$.

\begin{figure}[ht]\centering
	\includegraphics[width=9cm]{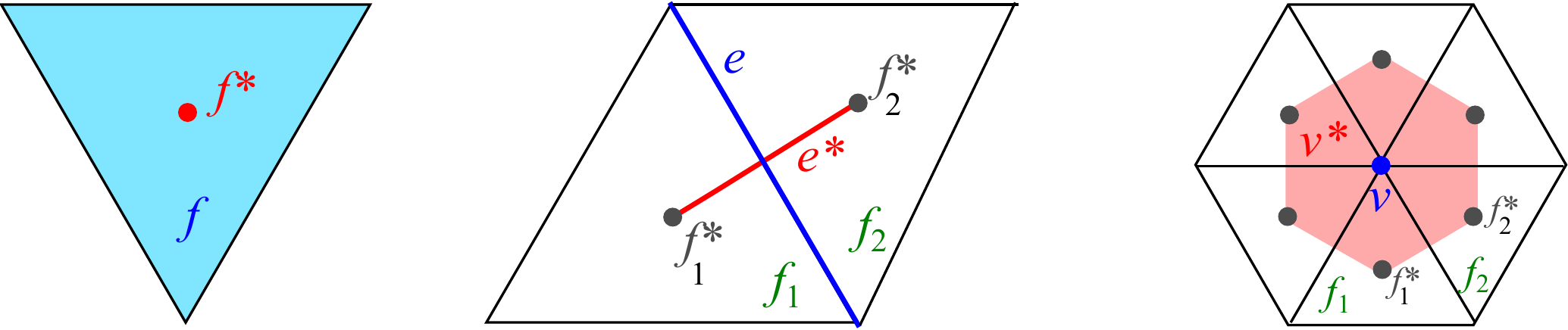}
	\caption{Primal simplices (in blue) of a triangle $f$ (left), an edge $e$ (middle) and a vertex $v$ (right) and their duals $f^*$, $e^*$, $v^*$ (in red) in a 2D mesh}
	\label{fig:dual}
\end{figure}

\begin{figure}[ht]\centering
	\includegraphics[width=3cm]{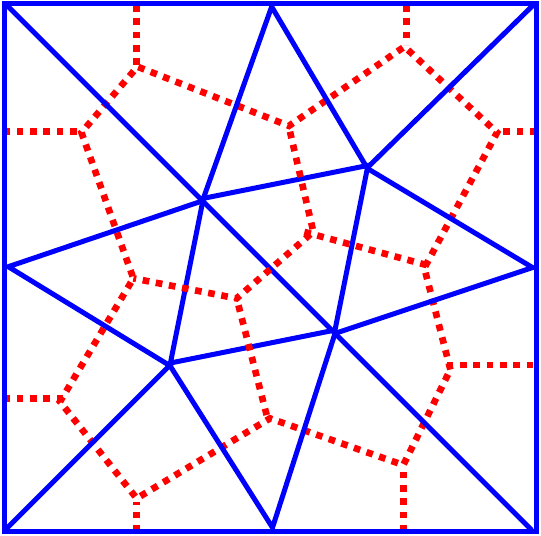}
	\caption{Sample 2D mesh (in blue) on a square and its circumcentric dual (in red)}
	\label{fig:maillage_dual}
\end{figure}

The discrete Hodge star operator should realize an isomorphism between $Ω^k(K)$ and $Ω^{n-k}(K)$. This is not possible on the same grid $K$ because, generally,
$$\operatorname{dim}Ω^k(K)=\operatorname{card}K_k\not=\operatorname{dim}Ω^{n-k}(K)=\operatorname{card}K_{n-k}.$$
For example, with the mesh on Figure \ref{mesh}, $n$ equals 2 and if $k=0$,
\[\operatorname{dim}Ω^k(K)=4,\quad\quad\operatorname{dim}Ω^{n-k}(K)=2.\]
In order to define a discrete Hodge star operator, a dual mesh $\star K$ is needed. This dual mesh must be constituted by cells such that there is a one-to-one correspondance between $k$-simplices of $K$ and $(n-k)$-dimensional cells of $\star K$. One possible dual mesh is the circumcentric dual. In a 2D mesh such as in Figure \ref{fig:dual}, the circumcentric dual is such that the dual of a triangle is its circumcenter; the dual of a primal edge is the edge connecting the circumcenters of two triangles which share this primal edge; and the dual of a vertex is the 2-cell formed by connecting the circumcenters of the primal triangles which share this vertex. Figure \ref{fig:maillage_dual} presents an example of a mesh on a square and its circumcenter-based dual. Note that the choice of circumcenters as duals of triangles implies the orthogonality between edges and their duals.

The boundary and discrete exterior derivative can be transposed to the dual mesh. The orientation on the primal mesh $K$ also induces an orientation on $\star K$ (see \cite{desbrun05})\footnote{To borrow the language of Bossavit \cite{bossavit05}, the (inner-)orientation of each primal simplex, as defined in section \ref{sec:discretization_domain}, constitutes the outer-orientation of its dual cell.}.

A dual $k-$form can be defined as a map which assigns a real number to each element of $\star K_k$,:
\begin{equation}
	ω\ :\ \begin{array}{ccc}\star K_k&\longrightarrow& ℝ\\σ^*&\longmapsto&⟨ω,σ^*⟩.\end{array}
	\label{dual_cochain}
\end{equation}
Computationally, a dual $k$-form is represented by an array $(⟨ω,σ_i^*⟩)_{σ_i∈K_{n-k}}$.
The space of $k$-forms is denoted $Ω^k(\star K)$. To avoid confusion, an element of $Ω^k(K)$ is qualified as primal $k$-form.

%Note that a simplex $σ∈K$ intersects its dual $σ^*$ at its circumcenter $c_{σ}$. This point is considered as the center of both $σ$ and $σ^*$.
The center (the circumcenter if the mesh is circumcentric) $c_{σ}$ is also considered as the center of its dual $σ^*$.
The degrees of fredoom (dof) of a primal $k$-form are located at the center of the $k$-simplices of $K$. The dof of a dual $k$-form are located at the center of the $k$-cells of $\star K$, which are also the center of the $(n-k)$-simplices of $K$. For instance, on an $n$-dimensionnal mesh with $n≥2$, both the dof of a primal $0$-form and the dof of a dual $n$-form are located at the vertices of $K$. The dof of a primal 1-form and those of a dual $(n-1)$-form are located at the center of edges of $K$. The dof of a primal 2-form and the dof of a dual $(n-2)$-form are located at the center of triangles.

A discrete Hodge operator, realizing an isomorphism between the spaces $Ω^k(K)$ and $Ω^{n-k}(\star K)$, can now defined as:
	$$\star :\ Ω^k(K)\longrightarrow Ω^{n-k}(\star K)$$
	with
\begin{equation}
	\cfrac{⟪\star ω,c^*⟫}{|c^*|}=\cfrac{⟪ω,c⟫}{|c|}, \quad\text{ for any }c\in K_k,
	\label{hodge}
\end{equation}
where $c^*\in K_{n-k}^*$ designates the dual of the $k-$simplex $c$. In equation (\ref{hodge}), $|c|$ is a measure of $c$, which is set to $1$ if $c$ is a point and to its Euclidean measure otherwise. Relation (\ref{hodge}) can be understood as follows: the discrete forms $ω$ and $\star ω$, normalized by the measure of the simplex/cell on which they are applied, represent the same density, but act in complementary-dimensional facets. Relation (\ref{hodge}) can also be understood as a discretization of equality (\ref{ortho}), $c$ and $c^*$ forming complementary geometric objects.

\begin{figure}
	\centering
	\includegraphics[width=6cm]{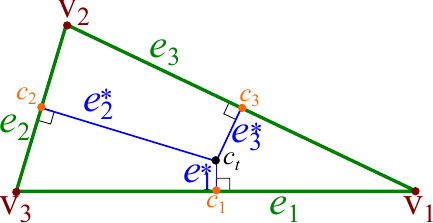}
	\caption{Primal simplices and their circumcentric dual cells in a 2D mesh composed of a single triangle. $c_i$ is the circumcenter of the primal edge $e_i$, for $i=1,2,3$, and $c_t$ is the circumcenter of the triangle. $e_i^*$ is perpendicular to $e_i$.}
	\label{triangle_circ}		
\end{figure}

Definition (\ref{hodge}) is a low-order approximation of the continuous Hodge star, which turns out to be exact in the particular case where $ω$ is piecewise constant and $c$ and $c^*$ are flat. It induces a representation of the discrete Hodge operator in each dimension as a diagonal matrix, having ratios of cell measures as entries. For example, in a one-triangle mesh as in Figure \ref{triangle_circ}, where $c_t$ and $c_i$ are the circumcenters of the triangle and the edges respectively, the Hodge matrix acting on primal 1-cochains is
\begin{equation}
	H_{diagonal}=
	\begin{bmatrix}
		\displaystyle\dfrac{|e_1^*|}{|e_1|}&0&0\\
		0&\displaystyle\dfrac{|e_2^*|}{|e_2|}&0\\
		0&0&\displaystyle\dfrac{|e_3^*|}{|e_3|}
	\end{bmatrix}. \label{hdiagonal}
\end{equation}
Due to relation (\ref{starstar}), the discrete Hodge matrix acting on dual 1-cochain is the inverse of matrix (\ref{hdiagonal}).

The diagonal structure of $H_{diagonal}$ and the induced computational efficiency makes the diagonal Hodge a popular choice of discrete Hodge operator. 
However, to be correctly defined, it requires a well-centered mesh. This means that each $k$-dimensional simplex of the mesh, with $k≥1$, must strictly contain its circumcenter. In 2D, a well-centered mesh can contain only strictly acute triangles. To understand the problem induced by the non well-centerdness, consider a 2D domain meshed with right triangles. The mesh is not well centered. Circumcenters lie on the boundary of the triangles and some dual edges degenerate into points (see Figure \ref{rectangle}, left, as an example), making matrix (\ref{hdiagonal}) singular. If the mesh contain obtuse triangles then some circumcenters are located outside the triangles. In this case, an extension, introducing a sign convention to the elementary volumes of the dual cells, has been proposed in \cite{mohamed18} to widen the range of applicability of the diagonal Hodge, but bringing no solution for the right-triangularized mesh for example (apart from altering the mesh). %
Moreover, the diagonal Hodge does not allow any choice than the circumcenters as duals of top-simplices. Indeed it has been shown that if $c_t$ and the $c_i$ are not the circumcenters, the diagonal formula (\ref{hdiagonal}) does not lead to a convergent Hodge operator \cite{mohamed16b}.

In the next section, we propose an alternative discrete Hodge operator.

\section{Analytically constructed discrete Hodge}

In the sequel, we restrict to the case of a bidimensional mesh in $ℝ^2$. We will fix a Cartesian orthonormal frame and denote $v_x$ and $v_y$ the components of a vector $\vec v$ in this frame. In the next subsection, we present the construction of a discrete Hodge on 1-forms. We require it to be exact on locally constant forms.

\subsection{Construction in a one-triangle mesh\label{sec:construction}}

\begin{figure}
	\centering
	\includegraphics[width=6cm]{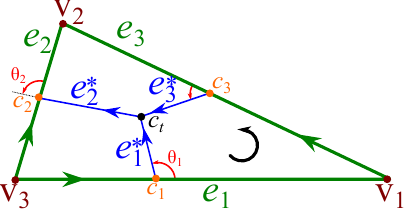}
	\caption{Primal simplices and their arbitrary-centered dual cells in a 2D mesh composed of a single triangle. The $c_i$'s and $c_t$ are respectively arbitrary interior points of the edges $e_i$'s and of the triangle. The triangle is oriented counterclockwise. Arrows indicate the orientations of the primal edges and the induced orientations of dual edges. The angles $θ_i$ defined in (\ref{angle}) are drawn in red.
	}
	\label{triangle}		
\end{figure}

To simplify, consider an oriented bidimensional mesh composed of a single triangle $T$, as in Figure \ref{triangle}. In this figure, the center $c_i$ of the edge $e_i$ is an arbitrary point lying on the edge. The center $c_t$ of the triangle $T$ is also an arbitrary point inside $T$. In the particular case where $c_i$ is the middle point of $e_i$ and $c_t$ is the barycenter (resp. circumcenter, incenter) of $T$, the dual mesh is called barycentric (resp. circumcentric, incentric).

Let us denote $\vec{e}_i$ and $\vec{e}_i^*$ the vectors representing the oriented edges $e_i$ and $e_i^*$ respectively, and \[\vec u_i=\dfrac{\ve_i}{\|\ve_i\|}\quad\quad\text{and}\quad\quad \vec u_i^*=\dfrac{\ve_i^*}{\|\ve_i^*\|}\] the corresponding unit vectors. 
These two vectors are related by a rotation:
\[ \vecteurd{u_{ix}^*}{u_{iy}^*}=\matriced{\cos\theta_i}{-\sin\theta_i}{\sin\theta_i}{\cos\theta_i}\vecteurd{u_{ix}}{u_{iy}} \]
where 
\begin{equation}
	\theta_i=\widehat{(\vec u_i,\vec u_i^*)}
	\label{angle}
\end{equation}
is the angle that they form (see Figure \ref{triangle}).

Consider a differential 1-form
\[ ω=aｄx+bｄy \]
where $a$ and $b$ are scalar functions of $x$ and $y$. Its image by the continuous Hodge star is (see equation (\ref{star2d})):
\[ \star ω=-bｄx+aｄy. \]
We denote $ω_i=⟨ω,e_i⟩$ and $ω^*_i=⟨\star ω,e_i^*⟩$ the values of the corresponding primal and dual cochains.
Assume that $\omega$ is constant inside the triangle. Then
\begin{eqnarray*}
	\omega_i^*&=&\int_{e^*_i}\!\star \omega\ =\ \star ω(u_i^*)\ |e_i^*|\ =(-bu_{ix}^*+au_{iy}^*)\ \ |e_i^*| \\
	   &=& \bigg((au_{ix}+bu_{iy})\ \sin\theta_i+ (-bu_{ix}+au_{iy})\ \cos\theta_i\bigg)|e_i^*|\\
	   &=& \bigg(ω(\vec u_i)\ \sin\theta_i+ ω(-\tsr J\vec u_i)\ \cos\theta_i\bigg)|e_i^*|,\\
	   &=& \bigg(÷{ω(\ve_i)}{|e_i|}\ \sin\theta_i+ ÷{ω(-\tsr J\ve_i)}{|\tsr J\ve_i|}\ \cos\theta_i\bigg)|e_i^*|,
 \end{eqnarray*}
$\tsr J$ being the skew-symmetric matrix \[\tsr J=\matriced {0}{-1}{1}{0}.\]
Knowing that $|\tsr J\ve_i|=|\ve_i|$, we have:
\begin{equation}
	\omega_i^*=\dfrac{|e_i^*|}{|e_i|}\omega_i\ \sin \theta_i\ \ +\ \ \dfrac{|e_i^*|}{|e_i|}\omega(-\tsr J\ve_i)\ \cos \theta_i.
	\label{swi}
\end{equation}

Let's take $i=1$. As soon as the triangle is not degenerate, there exists reals $a^2_1$ and $a_1^3$ such that 
\[ -\tsr J\ve_1=a^2_1 \ \ve_2+a^3_1\ \ve_3.\]
So,
\begin{equation}
	\omega_1^*=\dfrac{|e_1^*|}{|e_1|}\ \omega_1\ \sin \theta_1\ \ +\ \ \dfrac{|e_1^*|}{|e_1|}\ a_1^2\omega_2\ \cos \theta_1\ \ +\ \ \dfrac{|e_1^*|}{|e_1|}\ a_1^3\omega_3\ \cos \theta_1 .
	\label{a}
\end{equation}
Using similar arguments for $i=2$ and 3, we get:
\[\begin{bmatrix}
	\omega^*_1\\\\\omega^*_2\\\\\omega^*_3
\end{bmatrix}
=
\begin{bmatrix}
	\displaystyle\dfrac{|e_1^*|}{|e_1|}&0&0\\
	0&\displaystyle\dfrac{|e_2^*|}{|e_2|}&0\\
	0&0&\displaystyle\dfrac{|e_3^*|}{|e_3|}
\end{bmatrix}
\begin{bmatrix}
	\sin \theta_1 &a_1^2\cos\theta_1&a_1^3\cos\theta_1 \\\\
	a_2^1\cos\theta_2&\sin \theta_2 &a_2^3\cos\theta_2 \\\\
	a_3^1\cos\theta_3&a_3^2\cos\theta_3&\sin \theta_3 
\end{bmatrix}
\begin{bmatrix}
	\omega_1\\\\\omega_2\\\\\omega_3
\end{bmatrix}.
\]
This relation makes the discrete Hodge star operator appear as a product of two matrices. The first one contains ratios of dual and primal volumes. In case of circumcentric dual mesh, the second matrix reduces to the identity matrix and we retrieve the usual diagonal Hodge (\ref{hdiagonal}).

Of course, one can rewrite the angles $θ_i$ and the coefficients $a_i^j$ as functions of the $\ve_i$ and $\ve_i^*$. For example, 
\[ \cos\theta_i=\dfrac{\ve_i\cdotp\ve_i^*}{|e_i|\,|e_i^*|},\quad  \sin\theta_i=\dfrac{\ve_i\dtimes\ve_i^*}{|e_i|\,|e_i^*|} ,\quad\quad\theta_i\in[0,\pi]\]
where $\vec a·\vec b$ is the inner product of $\vec a$ and $\vec b$. The dotted cross product is defined as the third component of the usual cross product of the extension of $\vec a$ and $\vec b$ to $ℝ^3$:
\[\vec a\dtimes\vec b=a_xb_y-a_yb_x\]
which can also be considered as the real-valued 2D cross product.
Moreover, it is straightforward to show that, for $i≠j$,
\[ a^i_j=\dfrac{\ve_j\cdotp\ve_k}{\ve_i\dtimes\ve_k},\]
where $k≠i,j$. With these relations, the matrix of the Hodge operator acting on discrete primal 1-form is
	\[ 
	H =
	\begin{bmatrix}
		\di\dfrac{\ve_1\dtimes\ve_1^*}{|e_1|^2}&&\di \dfrac{\ve_1\cdotp\ve_1^*}{|e_1|^2}\dfrac{\ve_1\cdotp\ve_3}{\ve_2\dtimes\ve_3}&&\di \dfrac{\ve_1\cdotp\ve_1^*}{|e_1|}\dfrac{\ve_1\cdotp\ve_2}{\ve_3\dtimes\ve_2} \\[15pt]
		\di\dfrac{\ve_2\cdotp\ve_2^*}{|e_2|^2}\dfrac{\ve_2\cdotp\ve_3}{\ve_1\dtimes\ve_3}&&\di\dfrac{\ve_2\dtimes\ve_2^*}{|e_2|^2}&&\di \dfrac{\ve_2\cdotp\ve_2^*}{|e_2|^2} \dfrac{\ve_2\cdotp\ve_1}{\ve_3\dtimes\ve_1}\\[15pt]
		\di\dfrac{\ve_3\cdotp\ve_3^*}{|e_3|^2}\dfrac{\ve_3\cdotp\ve_2}{\ve_1\dtimes\ve_2}&&\di \dfrac{\ve_3\cdotp\ve_3^*}{|e_3|^2}\dfrac{\ve_3\cdotp\ve_1}{\ve_2\dtimes\ve_1}&&\di\dfrac{\ve_3\dtimes\ve_3^*}{|e_3|^2}
	\end{bmatrix}.
\]

\subsection{Examples and remarks\label{sec:examples}}

As illustration, consider the right triangle presented in Figure \ref{rectangle}. The circumcentric dual (left of the figure) is singular because the dual $e_3^*$ of $e_3$ is empty and its measure vanishes. This leads to a singular matrix $H$, as in the diagonal Hodge. With the barycentric dual (Figure \ref{rectangle}, right), $H$ reduces to the following non-singular matrix:
\begin{figure}
	\centering
	\includegraphics[width=40mm]{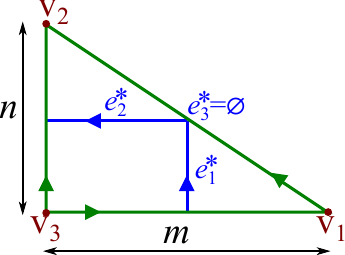}
	\hspace{1cm}
	\includegraphics[width=40mm]{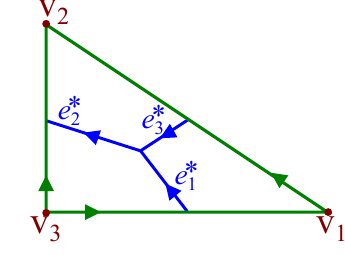}
	\caption{Right triangle with side lengths $m$ and $n$. Left: with a circumcentric dual mesh ($\e_3^*$ has a zero length). Right: with a barycentric dual mesh.}
	\label{rectangle}
\end{figure}
\[
H_{barycenter}=\begin{bmatrix}
	\di÷n{3m}&&\di÷{m}{6n}&&0
	\\[10pt]
	\di÷n{6m}&&\di÷m{3n}&&0
	\\[10pt]
	\di÷{n(m^2-n^2)}{6m(m^2+n^2)}&&\di÷{m(m^2-n^2)}{6n(m^2+n^2)}&&\di÷{mn}{3(m^2+n^2)}
\end{bmatrix}.
\]
In an unit right triangle ($m=n=1$), with a barycentric and an incentric dual meshes, $H$ reads:
%\begin{equation}
\[
	H_{barycenter}=÷16\begin{bmatrix}
		\di2&\di1&0
		\\
		\di1&\di2&0
		\\
		0&0&\di1
	\end{bmatrix} ,
	\quad\quad
	H_{incenter}=÷1{4+2√2}\begin{bmatrix}
		\di2&\di√2&0
		\\
		\di√2&\di2&0
		\\
		0&0&\di2
	\end{bmatrix}.
	\label{hodge_examples}
%\end{equation}
\]

The discrete Hodge operator for 0-forms and for 2-forms are composed of measure ratios as in the diagonal Hodge.

For a general mesh, composed of multiple triangles, a usual assembling permits to deduce the global Hodge matrix. This matrix is sparse. It takes a primal 1-cochain into a dual one. Again due to relation (\ref{starstar}), the Hodge operator acting on dual cochains can simply be represented by the inverse of the assembled matrix. However, a full inversion of this matrix is in general expensive. To avoid this cost, an element-wise inversion will be carried out in numerical tests when needed. Moreover, the element-wise inversion preserves the sparsity of the matrix, in contrast to a full inversion. This may significantly reduce the simulation cost for realistic problems. Despite this approximation, we will see in section \ref{sec:poisson} that the newly defined Hodge provides competitive numerical results.

As can be noted, the new discrete Hodge matrix is not necessarily symmetric. Yet, some authors claim that an Hogde matrix should be symmetric, since 
\begin{equation}
	\star ω\wedge θ =\star θ\wedge ω \label{starwedge}
\end{equation}
for two arbitrary 1-differential forms ω and θ, $\wedge$ being the wedge product (\cite{Auchmann06,hiptmair01}). Combined to a choice of volume form, this relation represents an inner product. However, relation (\ref{starwedge}) does not imply that the discrete Hodge has to be symmetric. Similarly, a symmetry of the matrix of the Hodge operator does not help to verify property (\ref{starwedge}). It is the combination of the discrete Hodge matrix with the discrete wedge operator between two discrete 1-forms which should be symmetric. It may also be advocated that a symmetric Hodge matrix would be more advantageous because it would result in a symmetric problem when discretizing a symmetric operator as in (\ref{starwedge}). This argument is however not convincing since the discrete wedge product may destroy the symmetry.

By definition, the symmetry of the Hodge operator matrix alone would mean, in a single-triangle mesh, that
\begin{equation}
	\begin{bmatrix}
		θ_1\\θ_2\\θ_3
	\end{bmatrix}·⦅H
	\begin{bmatrix}
		ω_1\\ω_2\\ω_3
	\end{bmatrix}⦆
	=⦅H
	\begin{bmatrix}
		θ_1\\θ_2\\θ_3
	\end{bmatrix}⦆·
	\begin{bmatrix}
		ω_1\\ω_2\\ω_3
	\end{bmatrix}
	\label{sym}
\end{equation}
for any couple $ω$ and $θ$ of primal 1-cochains, where the dot symbol $(\ ·\ )$ denotes the Euclidean scalar product of $ℝ^3$. Not only the scalar product between cochains has still to be given a sense but also there is no obvious reason to require condition (\ref{sym}).   

In fact, the symmetry requirement may be justified in a finite element approach as proposed by Hiptmair in \cite{hiptmair01} but not in DEC. Indeed, in a finite element framework, it is the mass matrix which is called Hodge operator. Yet, this mass matrix represents, not the Hodge operator, but inner products such as (\ref{starwedge}) integrated over elements. Note also that, as recognized in \cite{Auchmann06,Alotto04,cinalli04}, the symmetry restriction seems to be too strong in DEC context.

Which center gives more accurate results for non-constant 1-forms depends on the form. For instance, consider the differential 1-form 
\[ω=(x-y)(ｄx-ｄy).\] Table \ref{tab:error_hodge} lists the error of the discrete Hodge operator $H$ on ω, $H$ being either the barycentric or the incentric Hodge. The error is defined as the $\ell^2$-norm of the dual cochain $H(ω_K)-(\star ω)_{\star K}$. Two computations are carried out. The first one is on a mesh made up of the single unit triangle, having a mean edge length 1.1381. The second computation is on a square meshed with 722 right and isosceles triangles (similar to Figure \ref{right_and_acute_mesh}, right) where the mean edge length is $5.6965·10^{-2}$. Table \ref{tab:error_hodge}  shows that, with the considered form, the barycenter-based Hodge is more precise than the inceter-based one on the single-triangle mesh. The same conclusion can be drawn in the discretized square. However, in the latter case, the error difference is smaller. For the differential form \[ω=(x+y)(ｄx+ｄy),\] the situation is reversed. Indeed, Table \ref{tab:error_hodge} shows that for this form, the incentric Hodge is about 1.94 times as accurate as the barycentric one, as well on the single-triangle mesh as in the right-triangularized mesh.

\begin{table}[ht]
	\centering
	$\def\arraystretch{1.5}
	\begin{array}{|c|c|c|c|c|}
		\hline
		&\multicolumn{2}{c|}{\text{Single triangle}}&\multicolumn{2}{c|}{\text{Right mesh}}
		\\\cline{2-5}
		\text{Differential form}& \text{Barycentric}       &\text{Incentric}&\text{Barycentric}&\text{Incentric}
		\\\hline
		(x-y)(ｄx-ｄy)&0.2946&0.3232&1.5243·10^{-2}&1.5715·10^{-2}
		\\\hline
		(x+y)(ｄx+ｄy)&0.0589&0.0303&6.6882·10^{-4}&3.4424·10^{-4}
		\\\hline
	\end{array}
	$
	\caption{$\ell^2$-norm of the error of the discrete Hodge operators on the unit right triangle and on a right-triangularized square with 20 points in each direction}
	\label{tab:error_hodge}
\end{table}

This very simple test shows that it is not easy to predict which center will give the best results, in terms of precision. Moreover, in more concrete problems, there may be other constraints than the accuracy to account for. The freedom to choose the center makes it conceivable to carry out an optimization of the discrete Hodge operator.

Lastly, let us remark that, in the previous development, there is no restriction on the position of the centers of the simplices, as long as the dual mesh is not degenerate. For example, the center $c_t$ may be outside the triangle, or even outside the domain in the case of multiple-triangle mesh.

In the following section, numerical experiments on (linear) Poisson equation and applications to (non-linear) isothermal and anisothermal fluid flow problems are presented. We will in particular focus on the convergence with the circumcenter-, the barycentrer- and the incenter-based Hodges on different types of mesh.

\section{Numerical results\label{sec:numerical}}

We begin with numerical tests on Poisson problem.

\subsection{Poisson equation\label{sec:poisson}}

\begin{figure}[htb!]
\centering
	\includegraphics[scale=0.23]{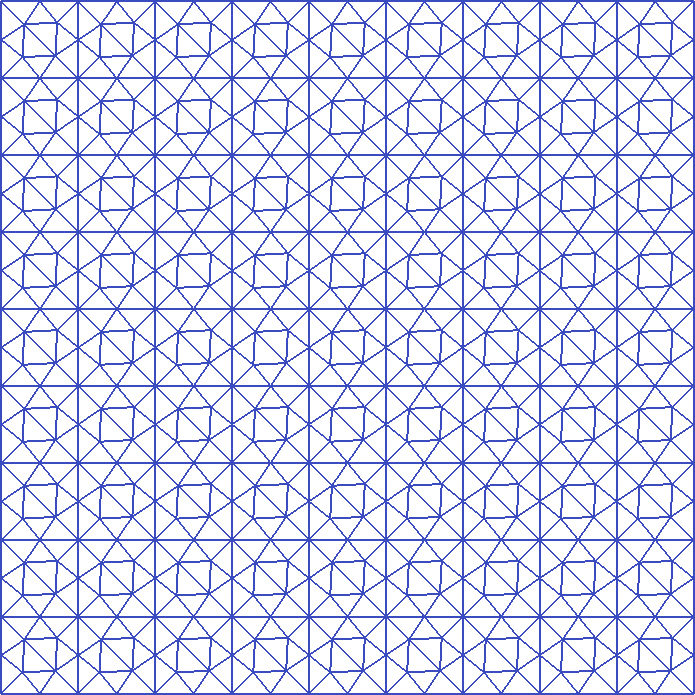}\quad\quad
	\includegraphics[scale=0.23]{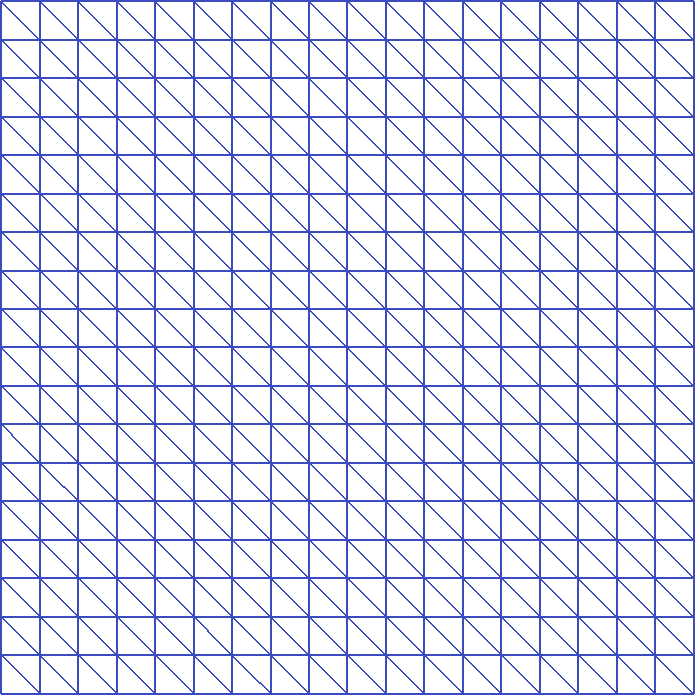}
	\caption{Typical acute and right triangulations of the unit square}
	\label{right_and_acute_mesh}		
\end{figure}

In this section, we deal with Poisson equation with Dirichlet boundary conditions in a flat bidimensional domain $M$:  
\begin{equation} \label{Poisson_Eq}
\begin{cases}\displaystyle
-\Delta u =f \quad \text{ in }  M,\\

u=g \quad\quad\text{\quad on }\partial M,
\end{cases}
\end{equation}
where $f$, $g$ and the unknown $u$ are scalar functions. In exterior calculus formulation, this equation reads
 \begin{equation} 
\label{eq:poisson_ext}
\begin{cases}\displaystyle
 \starｄ\star ｄu =f \quad \text{ in }  M,\\

u=g\quad\text{ \quad on }\partial M.
\end{cases}
\end{equation}
In the tests, $M$ is the unit square. Two meshes are considered. The first one is well-centered (see Figure \ref{right_and_acute_mesh}, left) and based on a pattern proposed in \cite{cassidi81}. It will be called ``acute mesh''. On this mesh, the usual diagonal Hodge and the new Hodge operator based on barycentric and on incentric duals will be compared. The second mesh, called ``right mesh'' herein, is more natural and is made of right and isosceles triangles (see Figure \ref{right_and_acute_mesh}, right). As the diagonal Hodge cannot be defined on it, the new Hodge operator with barycentric and incentric duals will be used.

The variable $u$ is placed, as cochain, on dual vertices $σ^*∈K^*_2$. This means that the unknown is the array $(⟨u,σ^*⟩)_{σ^*∈K^*_2}$. 
A consequence is that $ｄu$, in equation (\ref{eq:poisson_ext}), is discretized into a dual 1-cochain and the right-most $\star$ operator in that equation is then represeted by the inverse of the assembled Hodge matrix. As mentioned, the inversion is done element-wise numerically.

\begin{figure}
	\centering
	\begin{subfigure}{\figwidth}
		\includegraphics[width=\textwidth]{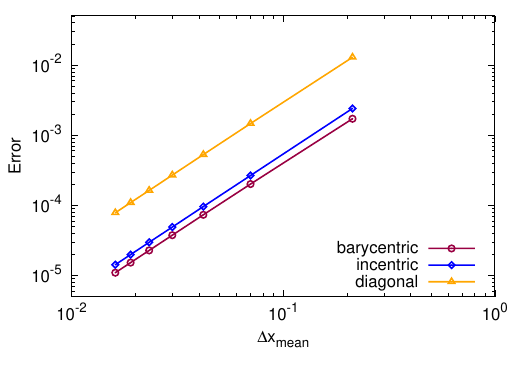}
		\caption{Acute mesh}\label{poisson_sin_acute}
	\end{subfigure}
	\begin{subfigure}{\figwidth}
		\includegraphics[width=\textwidth]{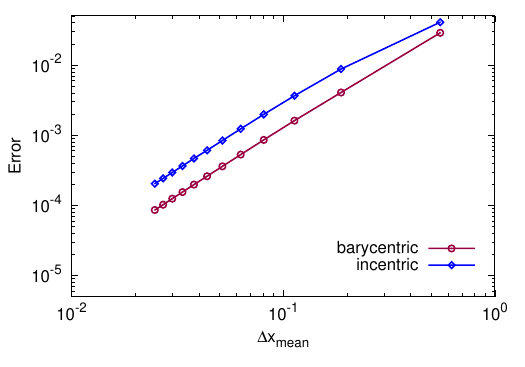}
		\caption{Right mesh}
	\end{subfigure}
	\caption{Poisson equation with with $u_{exact}=x^2+y^2$: error evolution}
	\label{fig:poisson_d_x2y2}
\end{figure}

As precision indicator, we consider the relative error 
\begin{equation}
	E=÷{‖u-u_{exact}‖_{K^*_2}}{‖u_{exact}‖_{K^*_2}}\label{relative_error_def}
\end{equation}
where $‖v‖_{K^*_2}$ designates a discrete norm of a dual 0-cochain $v$, defined as follows:
\begin{equation}
	‖v‖^2_{K^*_2}:=∑_{σ^*\in K^*_2}⟨v,σ^*⟩^2|\sigma^*|.\label{norm_def}
\end{equation}
An object-oriented Fortran code was designed for the simulations. A direct LU solver is selected for the resolution. In a first test, the right-hand side functions $f$ and $g$ are chosen such that the exact solution is the quadratic function 
\[u_{exact}=x^2+y^2.\]
The code is run untill time $t=T$ from which no more significant change is observed on the computed velocity field. Figure~\ref{fig:poisson_d_x2y2} shows the evolution of the relative error $E$ at $t=T$ with the mean edge length $Δx_{mean}$. As can be seen in Figure \ref{poisson_sin_acute}, barycentric and incentric dual meshes provide better results than the diagonal Hodge, despite the element-wise inversion needed in the discretized Laplace operator. Note that many triangles are close to right. However, the convergence rates are almost the same with the three types of dual mesh. The left part of Table \ref{tab:poisson_d_x2y2} reveals an almost quadratic convergence.

\begin{table}[ht]
	\centering
	\begin{tabular}[]{|l|c|c|}\hline
		&Acute mesh&Right mesh\\\hline
		Circumcentric dual&1.995&--\\\hline
		Barycentric dual&1.985&1.923\\\hline
		Incentric dual&1.992&1.921\\\hline
	\end{tabular}
	\caption{Poisson equation with with $u_{exact}=x^2+y^2$: convergence rate}
	\label{tab:poisson_d_x2y2}
\end{table}

With the right mesh, the results are less accurate but are still very acceptable as can be observed in Figure \ref{fig:poisson_d_x2y2}b.  With the barycentric dual, the relative error is about $7.55·10^{-5}$ when $Δx_{mean}\simeq 2.3·10^{-2}$ (whereas it was $2.28·10^{-5}$ with an acute mesh). The difference between the barycentric and the incentric duals are more pronounced. However, the convergence rates are similar, as can be seen in Table  \ref{tab:poisson_d_x2y2}, right.

\begin{figure}
	\centering
	\begin{subfigure}{\figwidth}
		\includegraphics[width=\textwidth]{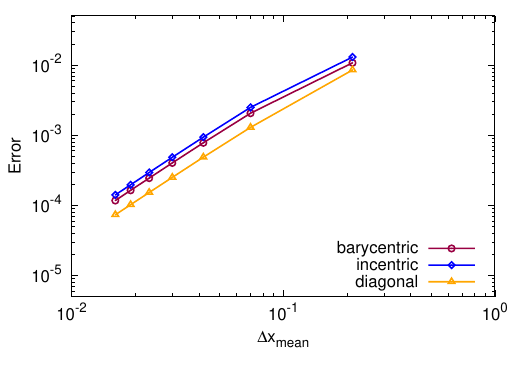}
		\caption{Acute mesh}
	\end{subfigure}
	\begin{subfigure}{\figwidth}
		\includegraphics[width=\textwidth]{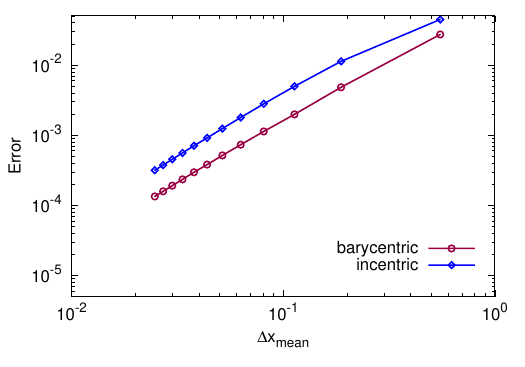}
		\caption{Right mesh}
	\end{subfigure}
	\caption{Poisson equation with $u_{exact}=\sin (πx)\,\sinh (πx)$: error evolution}
	\label{fig:poisson_d_sin}
\end{figure}

In a second test, the exact solution is \[u_{exact}=\sin (πx)\,\sinh (πy),\] such that the right-hand side $f$ vanishes. Figure \ref{fig:poisson_d_sin}a shows that, with the acute mesh, the three types of dual meshes present close precisions.
The barycentric solution is a slightly more accurate one with the right mesh whereas the incentric solution has clearly lost precision. They have however almost the same convergence rate, as can be seen in Table \ref{tab:poisson_d_sin}.

\begin{table}[ht]
	\centering
	\begin{tabular}[]{|l|c|c|}\hline
		&Acute mesh&Right mesh\\\hline
		Circumcentric dual&1.975&--\\\hline
		Barycentric dual&1.979&1.809\\\hline
		Incentric dual&1.982&1.840\\\hline
	\end{tabular}
	\caption{Poisson equation with $u_{exact}=\sin (πx)\,\sinh (πx)$: convergence rate}
	\label{tab:poisson_d_sin}
\end{table}

From these two experiments, it is obvious that when the mesh is well-centered, one cannot predict if the diagonal Hodge is more precise than other discrete Hodge operators. It depends on the solution. By contrast, the barycentric dual gave better accuracy than the incentric dual. %In fact, this last observation holds in each of the following numerical experiments.

In the previous tests, the equation is linear. In the next subsections, tests on non-linear systems in fluid dynamics field are presented.

\subsection{Incompressible fluid flow}

Consider an incompressible Newtonian fluid flow of density $\rho$ and kinematic viscosity $\nu$ in a bidimensional domain. The evolution of the velocity $u$ and the pressure $p$ is governed by the Navier-Stokes equations:
\begin{equation}
	\begin{cases}\displaystyle
		\dfrac{\partial u}{\partial t}+(u·∇)u + \dfrac {1}{\rho}\grad p-\nu\Delta u = 0,\\
		\dive u                                =0
	\end{cases}
	\label{ns}
\end{equation}
To ensure a divergence-free velocity, a stream function formulation is used. Moreover, to get rid of the pressure, a curl is applied on the momentum equation. 
The resulting equation is rewritten in exterior calculus formulation and discretized as in \cite{mohamed16a}, (see also equations (\ref{nsaext})). The discretized stream function is placed on primal vertices and, consequently, velocity is discretized into a dual 1-cochain.

\subsubsection{Poiseuille flow}

We consider a Dirichlet problem in the unit square with a Poiseuille flow as exact solution. The Reynolds number, based on the maximal velocity and the side length is 1. The accuracy of the stream function ψ and of the velocity $u$ is measured with the relative errors defined similarly as in (\ref{relative_error_def})-(\ref{norm_def}) but with the appropriate simplex/cell set instead of $K_2^*$.

\begin{figure}
	\centering
	\begin{subfigure}{\figwidth}
		\includegraphics[width=\textwidth]{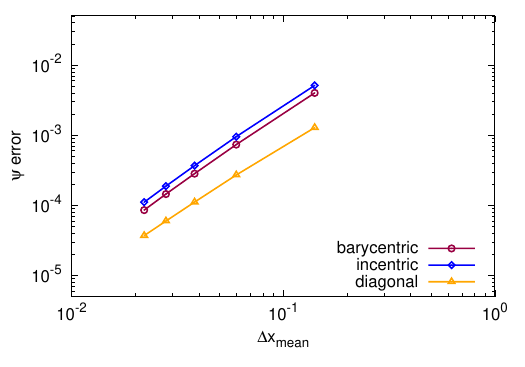}
		\caption{Acute mesh}
	\end{subfigure}
	\begin{subfigure}{\figwidth}
		\includegraphics[width=\textwidth]{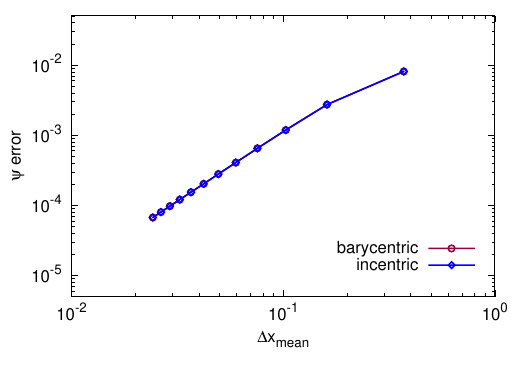}
		\caption{Right mesh}
	\end{subfigure}
	\caption{Poiseuille flow: error on $ψ$}
	\label{fig:poiseuille_psi}
\end{figure}

\begin{figure}
	\centering
	\begin{subfigure}{\figwidth}
		\includegraphics[width=\textwidth]{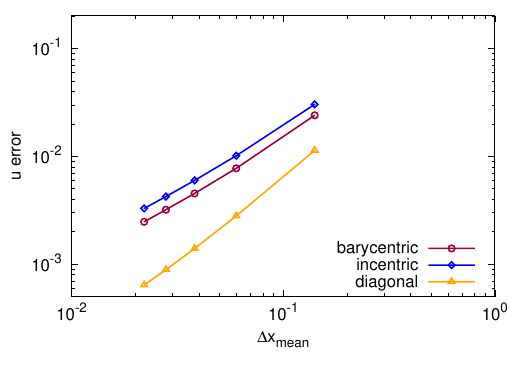}
		\caption{Acute mesh}\label{fig:poiseuille_u_acute}
	\end{subfigure}
	\begin{subfigure}{\figwidth}
		\includegraphics[width=\textwidth]{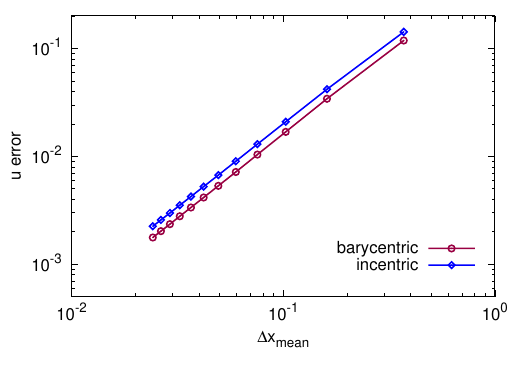}
		\caption{Right mesh}
	\end{subfigure}
	\caption{Poiseuille flow: error on $u$}
	\label{fig:poiseuille_u}
\end{figure}

The numerical initial flow is set to zero. The simulation is run until the flow stabilizes. The convergence of the principal variable ψ with the mean edge length is presented in Figure \ref{fig:poiseuille_psi}. It shows a higher precision of the diagonal Hodge in the acute triangulation in the given range of $Δx_{mean}$, but a slightly higher speed of convergence with the new Hodge with barycentric and incentric dual meshes. It is confirmed by the convergence rates in Table \ref{tab:poiseuille_psi_u}, second column. For the velocity however, the circumcentric dual has both a higher precision and a higher convergence speed, as can be observed in Figure \ref{fig:poiseuille_u_acute} and in the third column of Table \ref{tab:poiseuille_psi_u}. Note that a first order interpolation is used for the reconstruction of the approximate velocity field from the discrete 1-form $\star ｄψ$.

\begin{table}[ht]
	\centering
	\begin{tabular}{|c|c|c|c|c|}\hline
		&\multicolumn{2}{c|}{Acute mesh}&\multicolumn{2}{c|}{Right mesh}\\\cline{2-5}
		&Stream function&Velocity&Stream function&Velocity\\\hline
		Circumcentric&2.006 & 1.421&--&--\\
		Barycentric&2.184 & 1.106& 1.989 & 1.553\\
		Incentric&2.185 & 1.096& 1.989 & 1.539\\
		\hline
	\end{tabular}
	\caption{Poiseuille flow: convergence rates}
	\label{tab:poiseuille_psi_u}
\end{table}

In the case of the right triangulation, the barycentric and the incentric duals provide similar precisions, as well for the stream function as the velocity. For ψ, the errors are visually indistinguishable in Figure \ref{fig:poiseuille_psi}b. For the velocity, the convergence rate is about 1.5, which is higher than the convergence rate of the circumcentric dual in the acute triangulation case (see Table \ref{tab:poiseuille_psi_u}, last column).

\subsubsection{Pseudo Taylor-Green vortex}

We now simulate a steady Taylor-Green vortex corresponding to the vector field
\footnote{It is not the usual exponentially decaying Taylor-Green vortex but a time-independant vortex having the same shape as the Taylor-Green vortex} 
\begin{equation}u=
	\begin{pmatrix}
		-\cos x\,\sin y\\\sin x\cos y
	\end{pmatrix}
\end{equation}
in the domain $[-π,π]\times[-π,π]$. The kinematic viscosity and the density are set to 1.
As previously, the numerical initial flow is set to zero and the simulation is run until the flow stabilizes. 

\begin{figure}
	\centering
	\begin{subfigure}{\figwidth}
		\includegraphics[width=\textwidth]{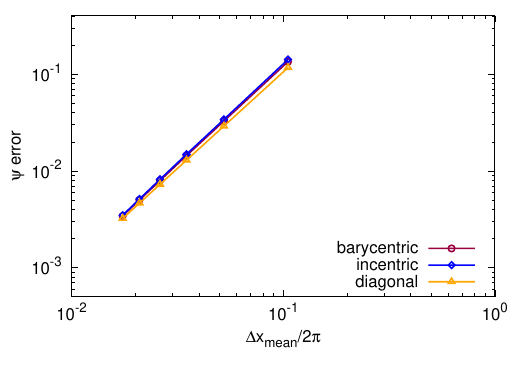}
		\caption{Acute mesh}
	\end{subfigure}
	\begin{subfigure}{\figwidth}
		\includegraphics[width=\textwidth]{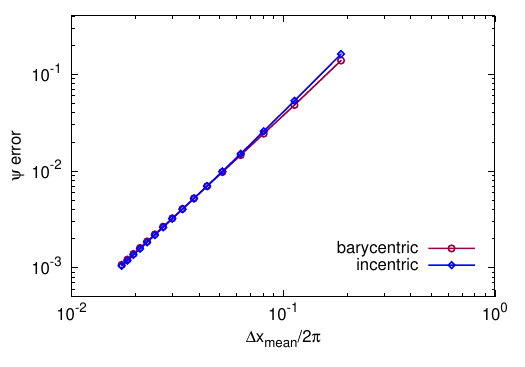}
		\caption{Right mesh}
	\end{subfigure}
	\caption{Taylor-Green vortex: error on $ψ$}
	\label{fig:taylor_psi}
\end{figure}

The error on ψ is plotted in Figure \ref{fig:taylor_psi} against the typical edge length. It shows that the decrease of the error is the same for the three types of dual mesh, as well in the acute mesh as in the right mesh. The convergence rate is close to 2.0 in all cases, as reported in Table \ref{tab:taylor_green_convergence_rates}.

\begin{table}[ht]
	\centering
	\begin{tabular}{|c|c|c|c|c|}\hline
		&\multicolumn{2}{c|}{Acute mesh}&\multicolumn{2}{c|}{Right mesh}\\\cline{2-5}
		&Stream function&Velocity&Stream function&Velocity\\\hline
		Circumcentric&1.996 & 1.187&--&--\\
		Barycentric&2.065 & 1.130& 2.018 & 1.735\\
		Incentric&2.088 & 1.118& 2.067 & 1.736\\
		\hline
	\end{tabular}
	\caption{Taylor-Green vortex: convergence rates}
	\label{tab:taylor_green_convergence_rates}
\end{table}

\begin{figure}
	\centering
	\begin{subfigure}{\figwidth}
		\includegraphics[width=\textwidth]{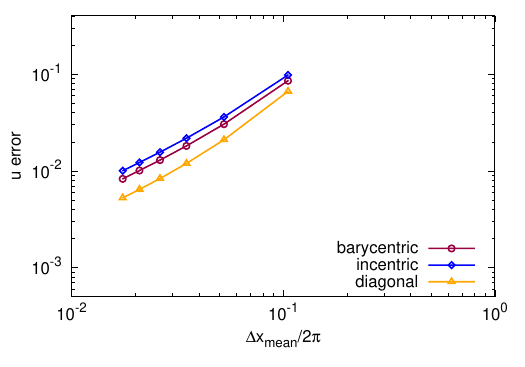}
		\caption{Acute mesh}
	\end{subfigure}
	\begin{subfigure}{\figwidth}
		\includegraphics[width=\textwidth]{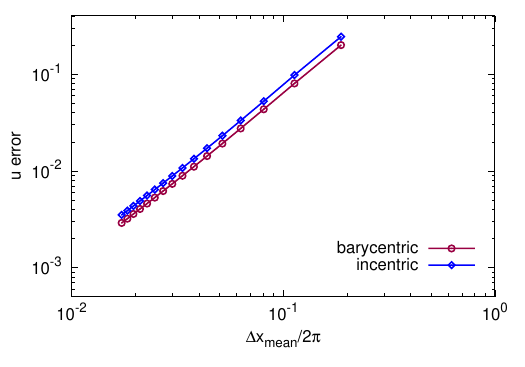}
		\caption{Right mesh}
	\end{subfigure}
	\caption{Taylor-Green vortex: error on $u$}
	\label{fig:taylor_u}
\end{figure}

For the velocity, differences can be observed between the acute and the right meshes. Indeed, it can be seen in Figure \ref{fig:taylor_u} that the decrease is faster in the right mesh. The convergence rate is about 1.74 (for both barycentric and incentric duals) while it is close to 1.14 in the acute mesh (for the three dual types), as can be seen in Table \ref{tab:taylor_green_convergence_rates}.

In the next test, we deal with an anisothermal fluid flow.

\subsection{Anisothermal fluid flow\label{sec:anisothermal}}

In the case of anisothermal fluid flow with thermal expansion coefficient $\beta$ and thermal diffusivity $\kappa$, and in the limit of the Boussinesq approximation, the evolution of the velocity, pressure and temperature $\theta$ is governed by the equations:
\begin{equation}
	\begin{cases}\displaystyle
		\dfrac{\partial u}{\partial t}+\dive(u\otimes u) + \dfrac {1}{\rho}\grad p-\nu\Delta u + \beta g \theta\e_y = 0,\\\\
		
		\dive u                                =0,\\\\
		\dfrac{\partial \theta}{\partial t}+\dive(u\theta)-\kappa\Delta\theta                  =0
	\end{cases}
	\label{nsa}
\end{equation}
where $-g\e_y$ is the gravity acceleration vector. Let $ω=u^\flat$. In exterior calculus framework, equation (\ref{nsa}) writes
\begin{equation}
	\begin{cases}\displaystyle
		\dfrac{\partial ω}{\partial t} +\iota_uｄω+ ÷1{ρ}ｄ\overline{p}+ν\starｄ\star ｄω+βgθｄy = 0,
		\\\\
		\star ｄ\starω=0,
		\\\\
		\dfrac{\partial \theta}{\partial t}+\starｄ\star (θω)-κ\starｄ\star ｄθ=0.
	\end{cases}
	\label{nsaext}
\end{equation}

In (\ref{nsaext}), $\iota$ is the interior product and $\overline{p}=p +÷12ρu^2$ is the dynamic pressure. The exterior derivative operator $ｄ$ is applied to the first equation of (\ref{nsaext}) to get rid of the dynamic pressure. A stream-function formulation is used. This means that the equations are solved in the stream function $ψ$ such that $\star ω=ｄψ$ instead of ω, and in the temperature θ. The time discretization scheme is similar to that in \cite{amses18}, page 55.

Assume that $\nu\neq \kappa$. We choose the following traveling wave solution as reference:
\begin{equation}
	\begin{cases}\displaystyle 
		u_x=u_1\e^{\lambda\xi/\nu}+\dfrac{\kappa^2\beta abg}{(c+w)^2(\kappa-\nu)}\theta_1\e^{\lambda\xi/\kappa}
		\\\\\displaystyle 
		u_y=\dfrac{w-au_x}b,
		\\\\\displaystyle 
		p=-\dfrac{\rho\beta bg\kappa\theta_1}{c+w}\e^{\lambda\xi/\kappa}
		\\\\\displaystyle 
		\theta=\theta_1\e^{\lambda\xi/\kappa}
	\end{cases}
	\label{travel1}
\end{equation}
where the traveling wave variable is
\begin{equation}
	\xi=ax+by+ct
	\label{var}
\end{equation}
and $a$, $b$, $c$, $w$, $u_1$, $λ$ and $θ_1$ are constants such that $b\neq 0$, $w≠-c$ and $λ=(c+w)/(a^2+b^2)$. 
The spacial domain is $\left[-\frac 12,\frac 12\right] \times\left[-\frac 12,\frac 12\right]$.  
The fluid is characterized by $\rho=1$, $\nu=0.2$, $\kappa=0.1$ and $\beta=1$.
The gravity acceleration is set to $g=10$. The values of the different constants in the reference solution are set to $a=b=1$, $c=-1$, $w=2$, $θ_1=\e^{-5}$ and $u_1=2\e^{-5}$. The Reynolds number based on the maximal value of $u_x$ at $t=0$ and the height $h=1$ of the domain is 4.18. The Rayleigh number based on $h$ and the temperature difference at $(÷12,0)$ and $(-÷12,0)$ at $t=0$ is 40.77.

The domain is discretized with a right triangulation (Figure \ref{right_and_acute_mesh}, right). The mean edge length is $1.894\cdot10^{-2}$. The numerical simulation is carried out until time $T$ such that $|\lambda c|T/\kappa=1$.

\begin{figure}
	\centering
	\includegraphics[width=\figwidth]{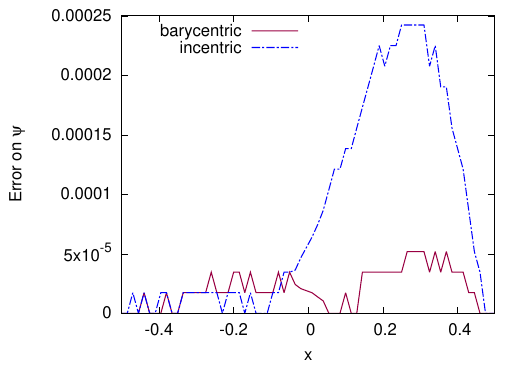}
	\caption{Traveling wave. Profile of the relative error on the stream function along $y=0$ and at $t=T$}
	\label{travel_psi_error}
\end{figure}

Figure \ref{travel_psi_error} reproduces the relative errors of the stream function along $x$-axis. Even if this figure presents some weird plateau due to lack of resolution (and some machine round-off), it clearly shows a good precision with both dual meshes. At worst, the error is 0.025 percent of the norm of the exact stream function along the axis. The mean error over the whole domain is still smaller. Indeed, as listed in Table \ref{tab:travel_error}, the overall (root mean square over the whole domain) error is only 0.008875 percent of the norm of the exact solution with the incentric dual and 3.3 times smaller with the barycentric one.

\begin{table}[ht]
	\centering
	\begin{tabular}{|c|c|c|c|c|c|}\hline
		Dual mesh&Stream function& Velocity&Temperature  \\\hline
		Barycentric&$2.651\cdot 10^{-5}$&$7.270\cdot10^{-5}$&$5.529\cdot10^{-3}$\\\hline
		Incentric&$8.875\cdot 10^{-5}$&$2.132\cdot10^{-4}$&$5.589\cdot10^{-3}$\\\hline
	\end{tabular}
	\caption{Traveling wave. Mean relative error}
	\label{tab:travel_error}
\end{table}

\begin{figure}
	\centering
	\includegraphics[width=\figwidth]{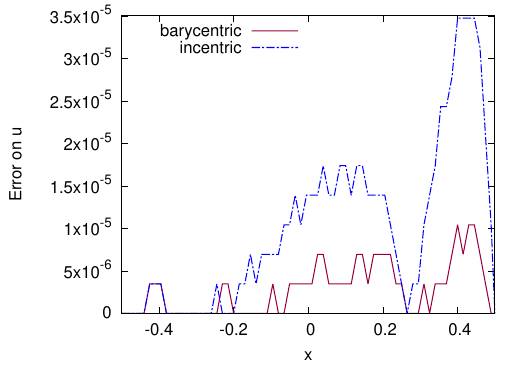}
	\caption{Traveling wave. Profile of the relative error on the horizontal velocity along $y=0$ and at $t=T$}
	\label{travel_u_error}
\end{figure}

Since the velocity is computed a posteriori from the stream function, the same conclusion can be drawn on the error. Indeed, the barycentric dual is more accurate  along the $x$-axis as shown in Figure  \ref{travel_u_error}. It is also more accurate in mean in the whole domain, as can be checked in Table \ref{tab:travel_error}.

\begin{figure}
	\centering
	\includegraphics[width=\figwidth]{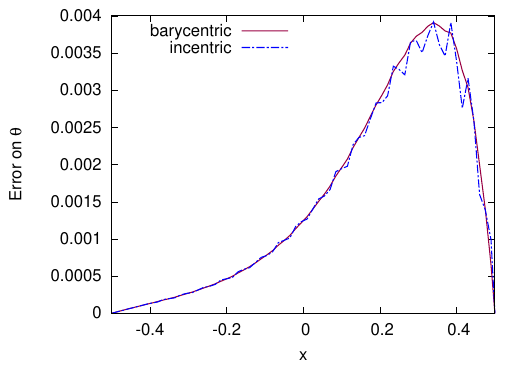}
	\caption{Traveling wave. Profile of the relative error on the temperature along $y=0$ and at $t=T$}
	\label{travel_theta_error}
\end{figure}

By contrast, for the temperature, the two dual meshes have a similar precision. The error profiles along $x$-axis, presented in Figure \ref{travel_theta_error}, have the same trend. The overall errors are almost equal as can be seen in the last column of Table \ref{tab:travel_error}. On average, the relative error is about 0.56 percent. These results reveals a good precision of the new discrete Hodge star on a heat transfer problem, with different choices of dual meshes. Convergence rates will be analyzed in the next section.

\subsection{Non-structured meshes\label{sec:nonstructured}}

To facilitate the reading of the article, only two types of primal mesh were considered in the above tests. The first ones were well-centered meshes such that the diagonal Hodge can be used. The second type of mesh was composed of right meshes. These meshes have structured patterns. In this following section, we show that the constructed Hodge gives also satisfactory results on various non-structured and non well-centered meshes. To simplify, only results on barycentric duals are presented.

We set $ν=κ$ and choose the following exact solution:
\begin{equation}
	\begin{cases}\displaystyle 
		u_x=\dfrac{\beta abg\theta_1}{(c+w)(a^2+b^2)}(x_0-\xi)\e^{\lambda\xi/\kappa}
		\\\\\displaystyle 
		u_y=\dfrac{w-au_1}b,
		\\\\\displaystyle 
		p=-\dfrac{\rho\beta bg\kappa\theta_1}{c+w}\e^{\lambda\xi/\kappa}
		\\\\\displaystyle 
		\theta=\theta_1\e^{\lambda\xi/\kappa}
	\end{cases}
	\label{travel2}
\end{equation}
where, as before, $ξ=ax+by+ct$ is the traveling wave variable.
For the simulation, the fluid characteristics are $\rho=1$, $g=10$, $\nu=\kappa=0.1$, $\beta=1$. We choose $a=b=1$, $c=-1$, $x_0=÷12$ and $w=0$. The spatial domain is $[0,1]\times[-\frac{1}{2},\frac{1}{2}]$.

\begin{figure}
	\centering
	\begin{tabular}{cc}
		\includegraphics[width=52mm]{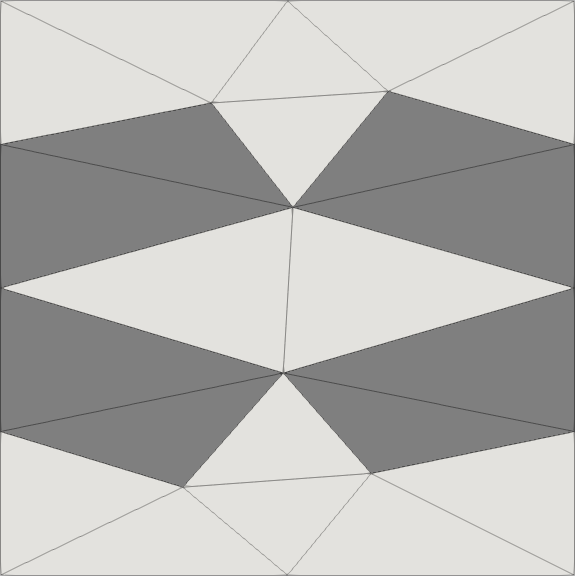}
		&
		\includegraphics[width=52mm]{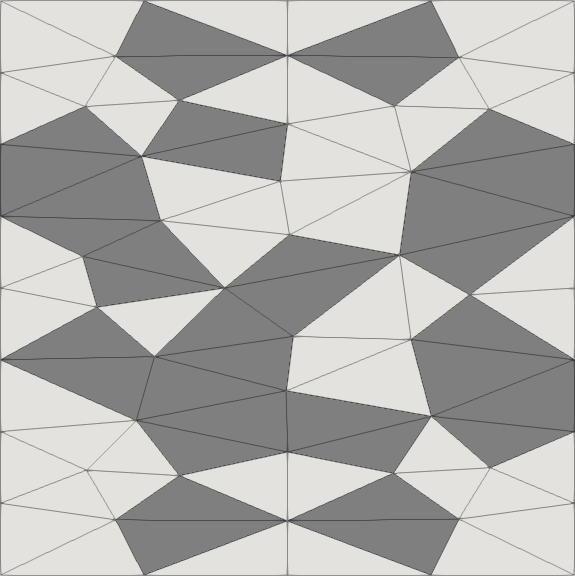}
		\\
		(a) $Δx_{mean}=0.35620$ & (b) $Δx_{mean}=0.18510$
		\\\\
		\includegraphics[width=52mm]{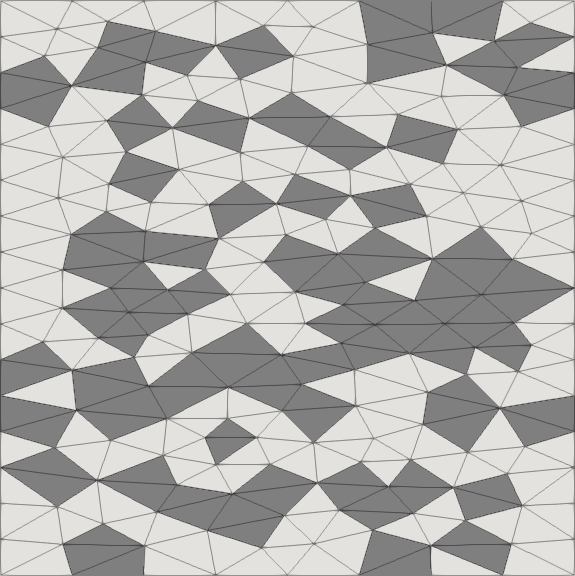}
		&
		\includegraphics[width=52mm]{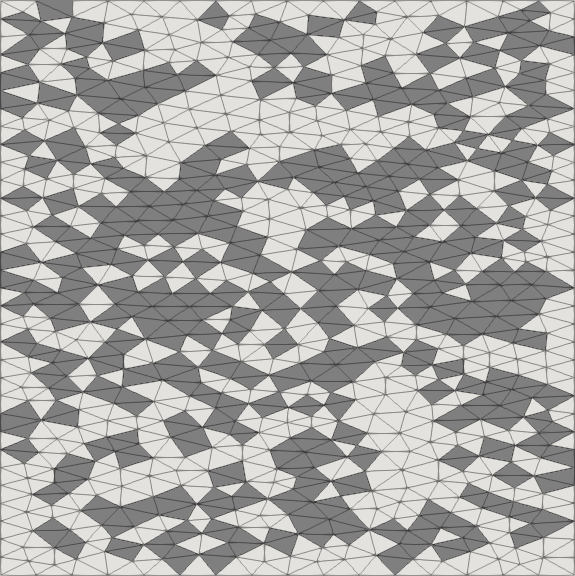}
		\\
		(c) $Δx_{mean}=0.09093$ & (d) $Δx_{mean}=0.04593$
		\\\\
		\multicolumn{2}{c}{\includegraphics[width=52mm]{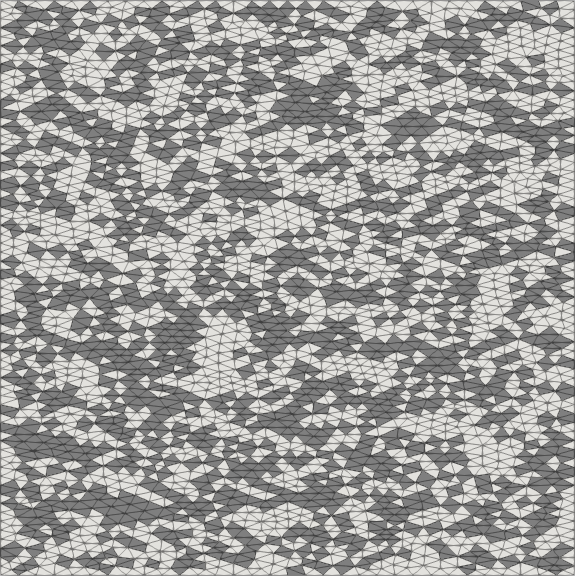}}
		\\
		\multicolumn{2}{c}{(e) $Δx_{mean}=0.02285$}
	\end{tabular}
	\caption{Unstructured meshes with respectively 36.36\%, 40.69\%, 39.41\%, 43.46\% and 44.20\% of non-Delaunay triangles}
	\label{fig:compressed_mesh}
\end{figure}

A first test is carried out on the five meshes shown in Figure  \ref{fig:compressed_mesh}. These meshes are increasingly fine, but are all unstructured, non well-centered and non-Delaunay. They have approximately 40 percent of non-Delaunay triangles. Here, a triangle is called non-Delaunay if its circumcircle contains other nodes than its vertices. They are darkened in the figures. Note that a non-Delaunay mesh is always non well-centered \cite{vanderzee10}.

\begin{figure}
	\centering
	\includegraphics[width=\figwidth]{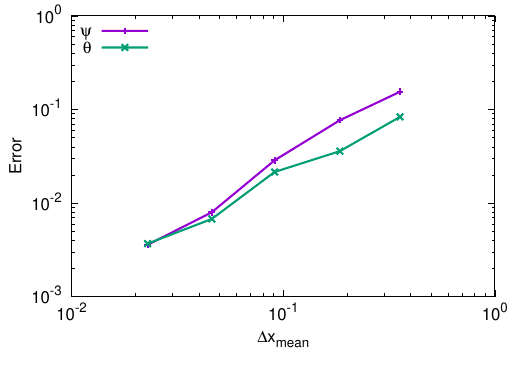}
	\caption{Unstructured meshes. Convergence of the stream function and the temperature}
	\label{fig:compressed_convergence}
\end{figure}

Figure \ref{fig:compressed_convergence} shows the relative overall errors on the stream function and on the temperature with the five meshes. As can be noted, the error decreases with the mean edge length. The decrease rates, which are reported in Table \ref{tab:compressed_rate}, are about 1.3690 for $ψ$ and 1.1430 for $θ$.

\begin{table}[ht]
	\centering
	\begin{tabular}{|c|c|c|}\hline
		    		&Stream function&Temperature  \\\hline
		    Convergence rate	&1.3690	&1.1430\\\hline
	\end{tabular}
	\caption{Unstructured meshes. Convergence rates of the stream function and the temperature}
	\label{tab:compressed_rate}
\end{table}

\begin{figure}
	\centering
	\includegraphics[width=85mm]{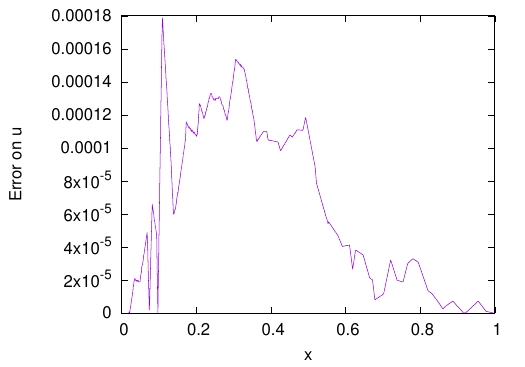}
	\includegraphics[width=85mm]{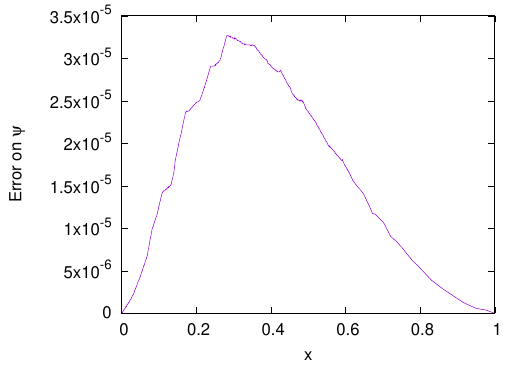}
	\includegraphics[width=85mm]{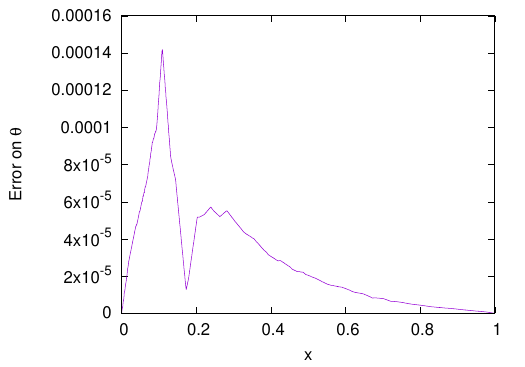}
	\caption{Relative error on mesh (e) of Table \ref{fig:compressed_mesh}}
	\label{fig:compressed_error}
\end{figure}

The relative error profiles on the horizontal velocity, the stream function and the temperature along the $x$-axis are plotted on Figure \ref{fig:compressed_error}. For simplicity, only the results with the finest mesh, that is mesh (e) of Figure \ref{fig:compressed_mesh}, are shown. The mean edge length in this mesh is $Δx_{mean}=2.285·10^{-2}$. Figure \ref{fig:compressed_error} shows that the errors along the axis remains very small compared to the norm of the exact solutions. The overall errors are recorded in Table \ref{tab:compressed_error}. As can be stated, the error on the velocity is about 2.316 percent of the norm of the exact solution, whereas the error on the stream function and on the temperature are less than 0.4 percent.

\begin{table}[ht]
	\centering
	\begin{tabular}{|c|c|c|c|c|c|}\hline
			       &Stream function        &Temperature         &Velocity                \\\hline
		Relative error &$3.581\cdot 10^{-3}$	&$3.682\cdot10^{-3}$&$2.316\cdot10^{-2}$	\\\hline
	\end{tabular}
	\caption{Relative errors on mesh (e) of Table \ref{fig:compressed_mesh}}
	\label{tab:compressed_error}
\end{table}

The meshes used for the previous results are unstructured and highly non-Delaunay. However, they are rather regular in the sense that the triangles have comparable areas and the edges have comparable lengths. In the last test, an analysis on non-Delaunay meshes with some very-badly-shaped triangles are carried out. These meshes are obtained from Delaunay meshes generated by Gmsh \cite{guezaine09}, of which some random vertices are moved, in order to obtain a prescribed ratio of non-Delaunay triangles. Contrarily to the distortion procedure in \cite{mohamed18}, the non-Delaunay triangles are not necessarily pairwise.

\begin{figure}%[h]
	\centering
	\begin{tabular}{cc}
	\includegraphics[width=60mm]{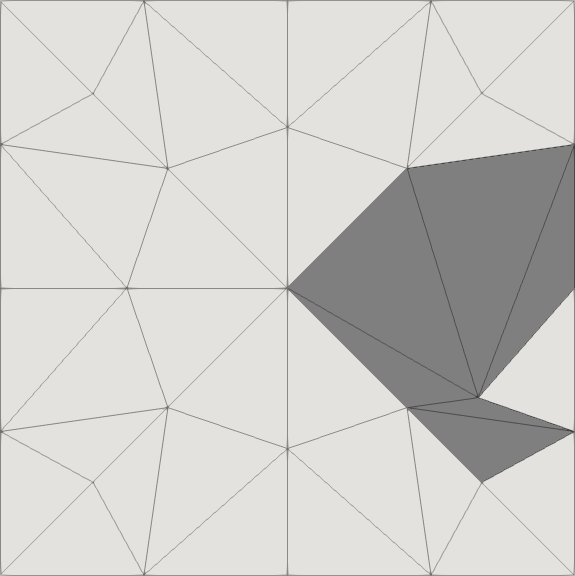}
		&
	\includegraphics[width=60mm]{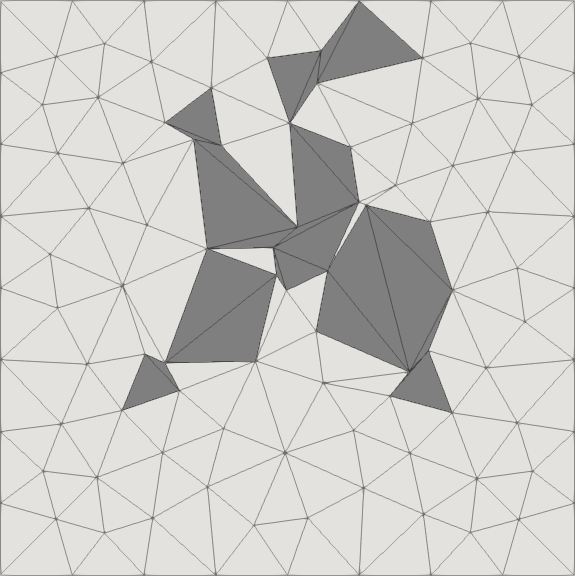}
		\\
		(a) $Δx_{mean}=0.2546$ & (b) $Δx_{mean}=0.1167$
		\\\\
	\includegraphics[width=60mm]{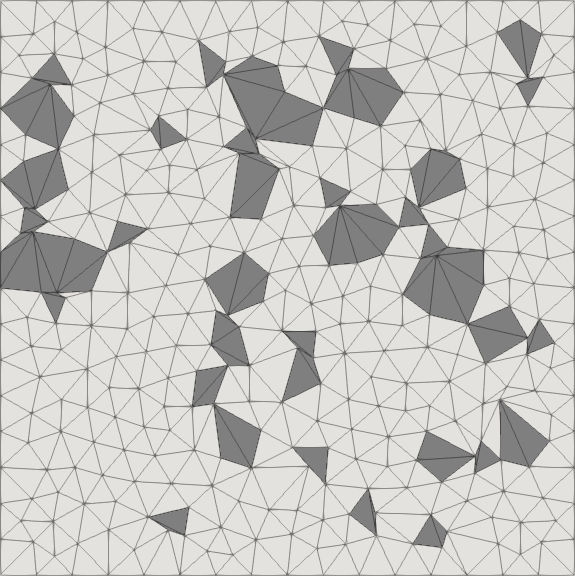}
		&
	\includegraphics[width=60mm]{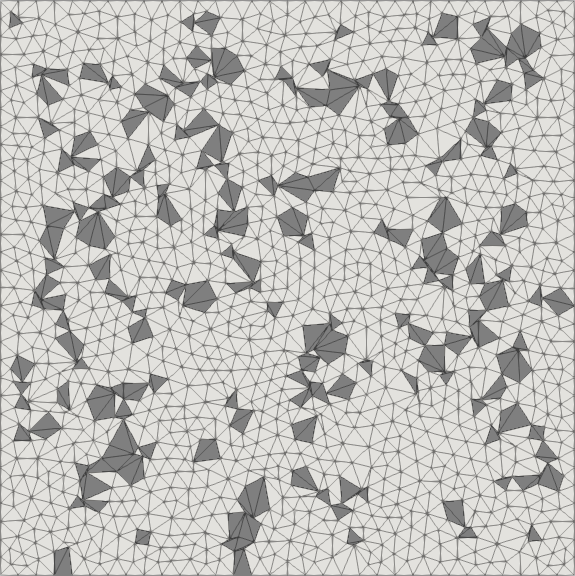}
		\\
		(c) $Δx_{mean}=0.0609$ & (d) $Δx_{mean}=0.0312$
	\end{tabular}
	\caption{Set of meshes with 15\% of non-Delaunay triangles}
	\label{fig:random15_mesh}
\end{figure}

\begin{figure}%[h]
	\centering
	\begin{tabular}{cc}
	\includegraphics[width=60mm]{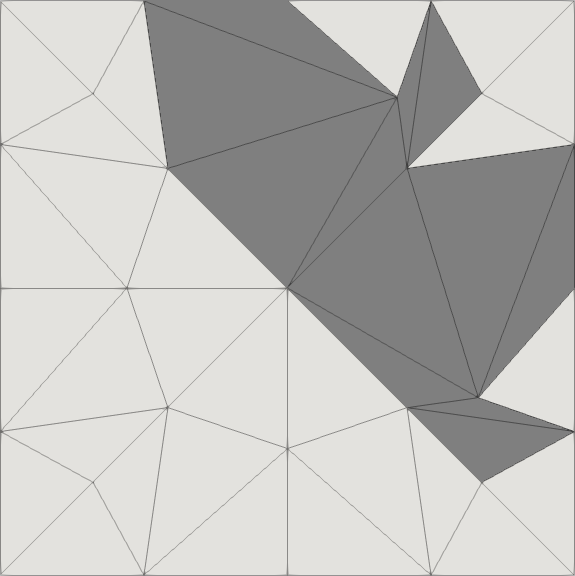}
		&
	\includegraphics[width=60mm]{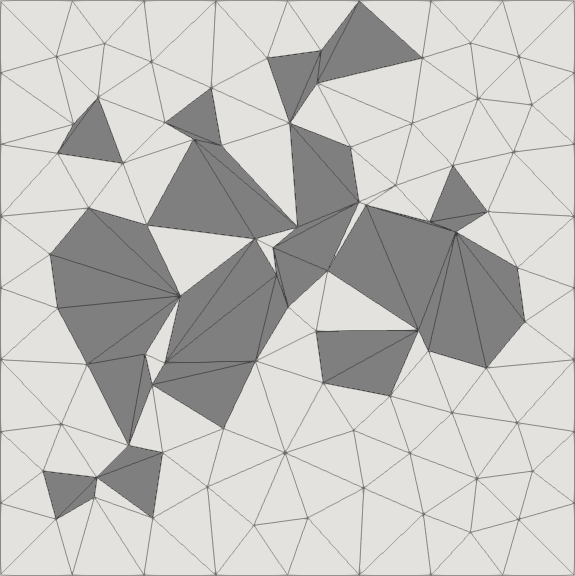}
		\\
		(a) $Δx_{mean}=0.2556$ & (b) $Δx_{mean}=0.1186$
		\\\\
	\includegraphics[width=60mm]{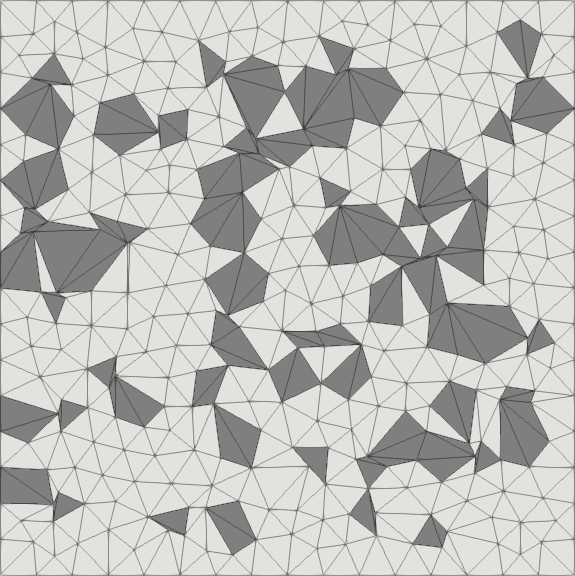}
		&
	\includegraphics[width=60mm]{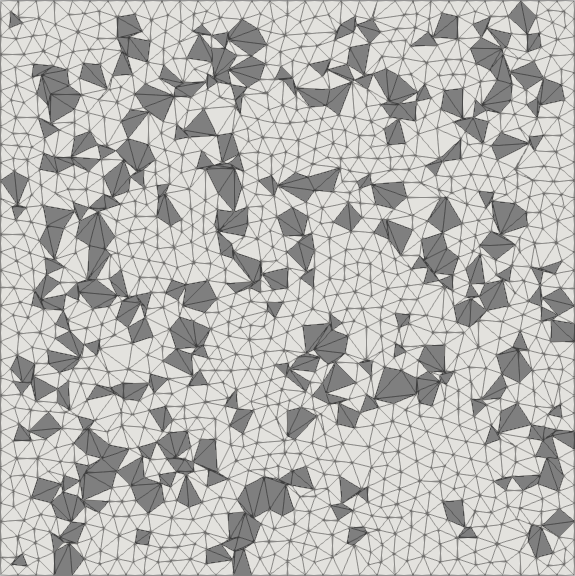}
		\\
		(c) $Δx_{mean}=0.0618$ & (d) $Δx_{mean}=0.0307$
	\end{tabular}
	\caption{Set of meshes with 25\% of non-Delaunay triangles}
	\label{fig:random25_mesh}
\end{figure}

\begin{figure}%[h]
	\centering
	\begin{tabular}{cc}
	\includegraphics[width=60mm]{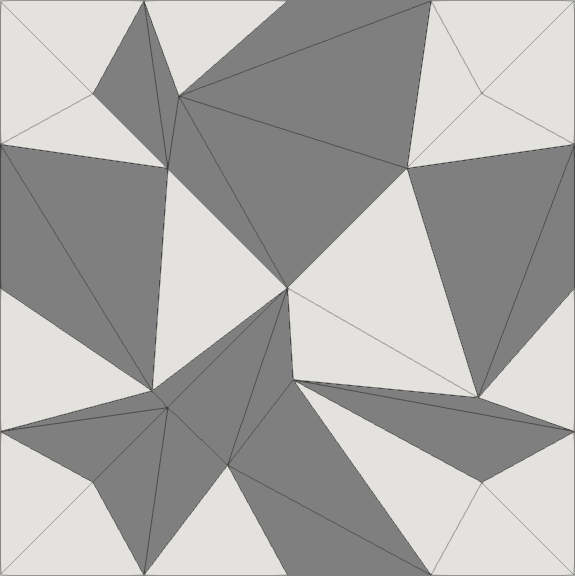}
		&
	\includegraphics[width=60mm]{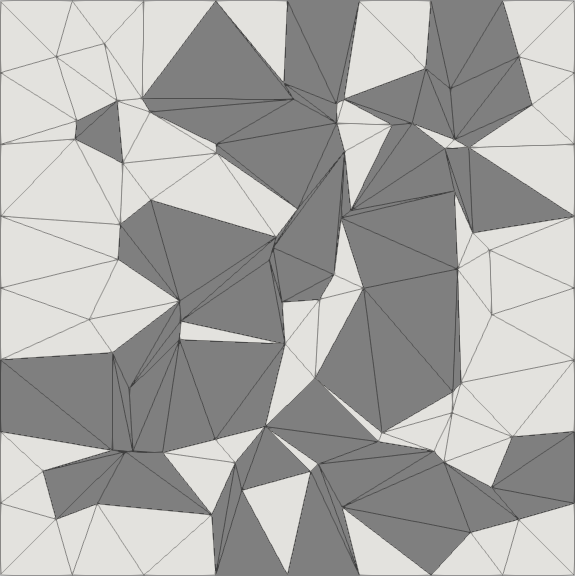}
		\\
		(a) $Δx_{mean}=0.2682$ & (b) $Δx_{mean}=0.1265$
		\\\\
	\includegraphics[width=60mm]{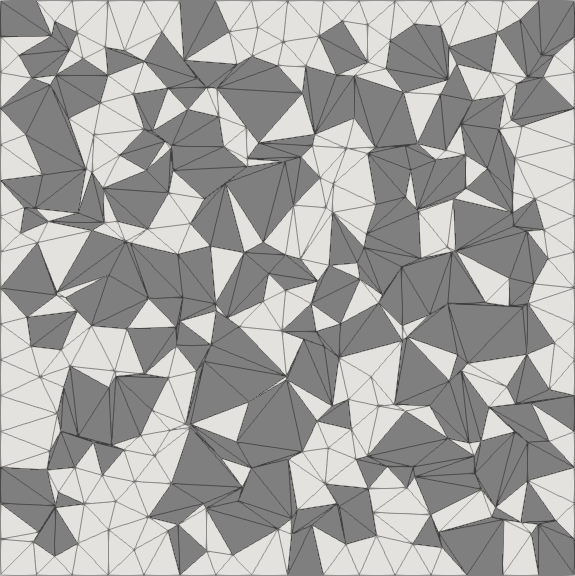}
		&
	\includegraphics[width=60mm]{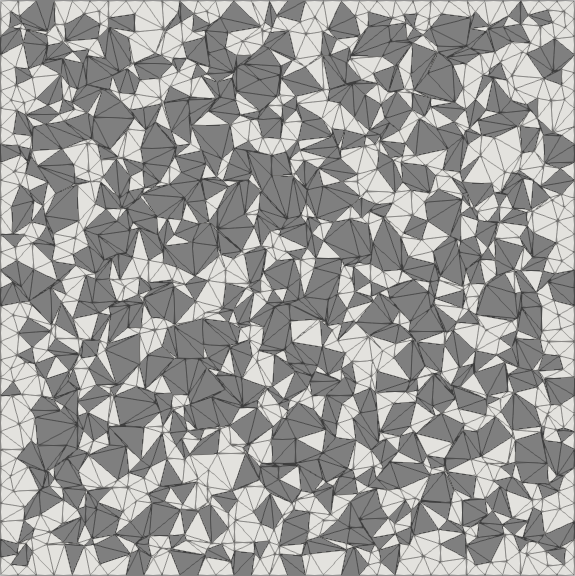}
		\\
		(c) $Δx_{mean}=0.0650$ & (d) $Δx_{mean}=0.0328$
	\end{tabular}
	\caption{Set of meshes with 50\% of non-Delaunay triangles}
	\label{fig:random50_mesh}
\end{figure}

Three sets of four increasignly fine meshes are used. In each set, the meshes count respectively 40, 184, 676 and 2658 triangles. The four meshes in the first set comprise approximately 15 percent of non-Delaunay triangles. They are presented in Figure \ref{fig:random15_mesh}. In the second set, the ratio of non-Delaunay triangles is about 25 percent, whereas in the last set, 50 percent of the triangles are non-Delaunay. The meshes in the second set are plotted in Figure \ref{fig:random25_mesh} and the third set can be seen in Figure \ref{fig:random50_mesh}. The mean edge lengths in the three sets range from $2.546·10^{-1}$ to $3.066·10^{-2}$. It can be noticed in Figures \ref{fig:random15_mesh} to \ref{fig:random50_mesh} that the triangles have irregular shapes. Many of them have an angle close to 180 degrees and the area of the triangles varies very significantly. For instance, the minimum triangle area of mesh (d) in Figure \ref{fig:random25_mesh} is $1.962·10^{-7}$ whereas the maximum is $2.228·10^{-3}$, that is more than $10^4$ times higher. 
%\begin{figure}
	%\centering
	%\begin{tabular}[]{|c|cc|cc|cc|cc|}
		%\hline
		%Percentage of&\multicolumn{2}{c|}{Mesh (a)}&\multicolumn{2}{c|}{Mesh (b)}&\multicolumn{2}{c|}{Mesh (c)}&\multicolumn{2}{c|}{Mesh (d)}\\
		%Non-Delaunay&Mean&Max&Mean&Max&Mean&Max&Mean&Max\\
		%\hline
		%15&ié&ii&uu\\
		%25&ié&ii&uu\\
		%50&ié&ii&uu
	%\end{tabular}
	%\caption{<+caption text+>}
	%\label{fig:<+label+>}
%\end{figure}<++>

\begin{figure}%[h]
	\centering
	\includegraphics[width=84mm]{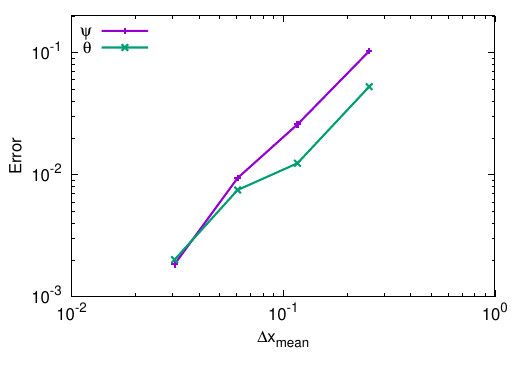}\\
	\includegraphics[width=84mm]{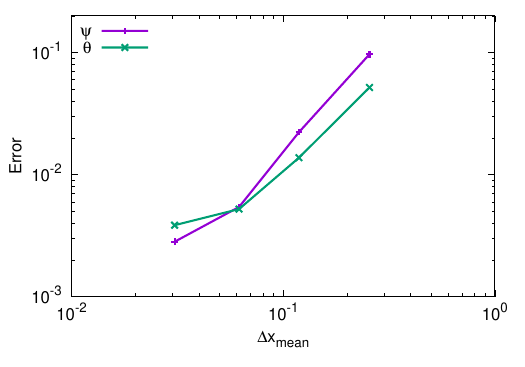}\\
	\includegraphics[width=84mm]{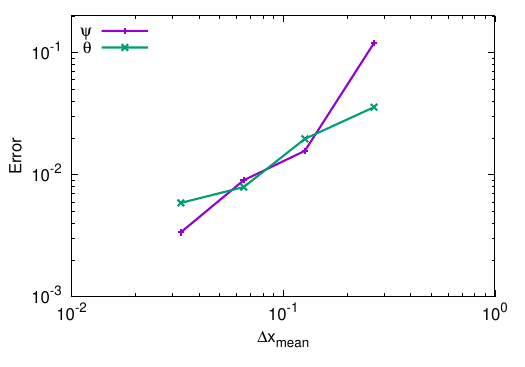}
	\caption{Convergence of the stream function and of the temperature. From top to bottom: with the first set (Figure \ref{fig:random15_mesh}), the second set (Figure \ref{fig:random25_mesh}) and the third set (Figure \ref{fig:random50_mesh}) of meshes}
	\label{fig:random_convergence}
\end{figure}

For all of these meshes, the computation runs without any particular complication. The errors decrease with the mean edge length. Figure \ref{fig:random_convergence} reports the evolution of the error on the stream function and on the temperature with the three sets of meshes. They both show a good convergence. For the first set of meshes, the convergence is close to second order for ψ and is at a rate of 1.5159 for θ, as listed on Table~\ref{tab:random_rate}. For the 25\% non-Delaunay meshes, the convergence rates are slightly smaller but remain satisfactory. They are 1.6729 for ψ and 1.2154 for θ. For the set of 50\% non-Delaunay meshes, ψ has a similar convergence rate as in the second set. For the temperature, the result is less satisfying since the convergence rate is smaller than 1.

\begin{table}[ht]
	\centering
	\begin{tabular}{|c|c|c|}\hline
		    		&Stream function&Temperature  \\\hline
		    %Compressed	&1.3690	&1.1430\\\hline
		    15\% non-Delaunay meshes &1.9005	&1.5159\\\hline
		    25\% non-Delaunay meshes &1.6729	&1.2154\\\hline
		    50\% non-Delaunay meshes &1.6591 	&0.8660\\\hline
	\end{tabular}
	\caption{Convergence rate}
	\label{tab:random_rate}
\end{table}

\begin{figure}%[h]
	\centering
	\includegraphics[width=\figwidth]{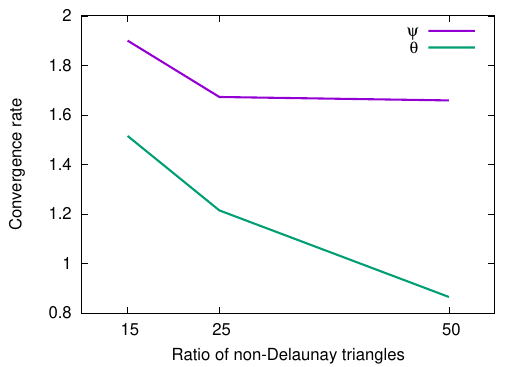}
	\caption{Evolution of the convergence rate with the ratio of non-Delaunay triangles}
	\label{fig:random_rate}
\end{figure}

The values in Table \ref{tab:random_rate} are plotted in Figure \ref{fig:random_rate} which presents visually the evolution of the convergence rate with the mesh quality. This figure shows that the convergence rate decreases less rapidely than the rate of non-Delaunay triangles.

\subsection{Fluid flow on a surface\label{sec:surface}}

At last, a numerical experiment on a non-flat surface in $ℝ^3$ is carried out. Before presenting the results, it is worth to warn the reader that the correct equations governing a fluid flow on a surface are still an open question. Various assumptions can be found in literature and they lead to various formulations. For our test, we choose the model proposed in \cite{nitschke17} where the surface Navier-Stokes equations read
\begin{equation}
	\begin{cases}\displaystyle
		\dfrac{\partial u}{\partial t}+(u·∇)u + \dfrac {1}{\rho}\grad p-\nu\Delta u - 2γu=0,\\\\
		\dive u                                =0
	\end{cases}
	\label{ns_surface}
\end{equation}
and $γ$ is the Gaussian curvature.

The equations are discretized similarly as in previous sections. The curvature is spatially discretized into a 0-cochain and the term $2γu$ is explicitely discretized in time. 

We consider the time development of an initial vortex on an ellipsoid with axes $1$, $1$ and $3$. The initial stream function is $ψ=y+0.1z$. The Reynolds number, based on the mean of the three axes and a unit velocity is 10. 

\begin{figure}%[h]
	\centering
	\includegraphics[width=7cm]{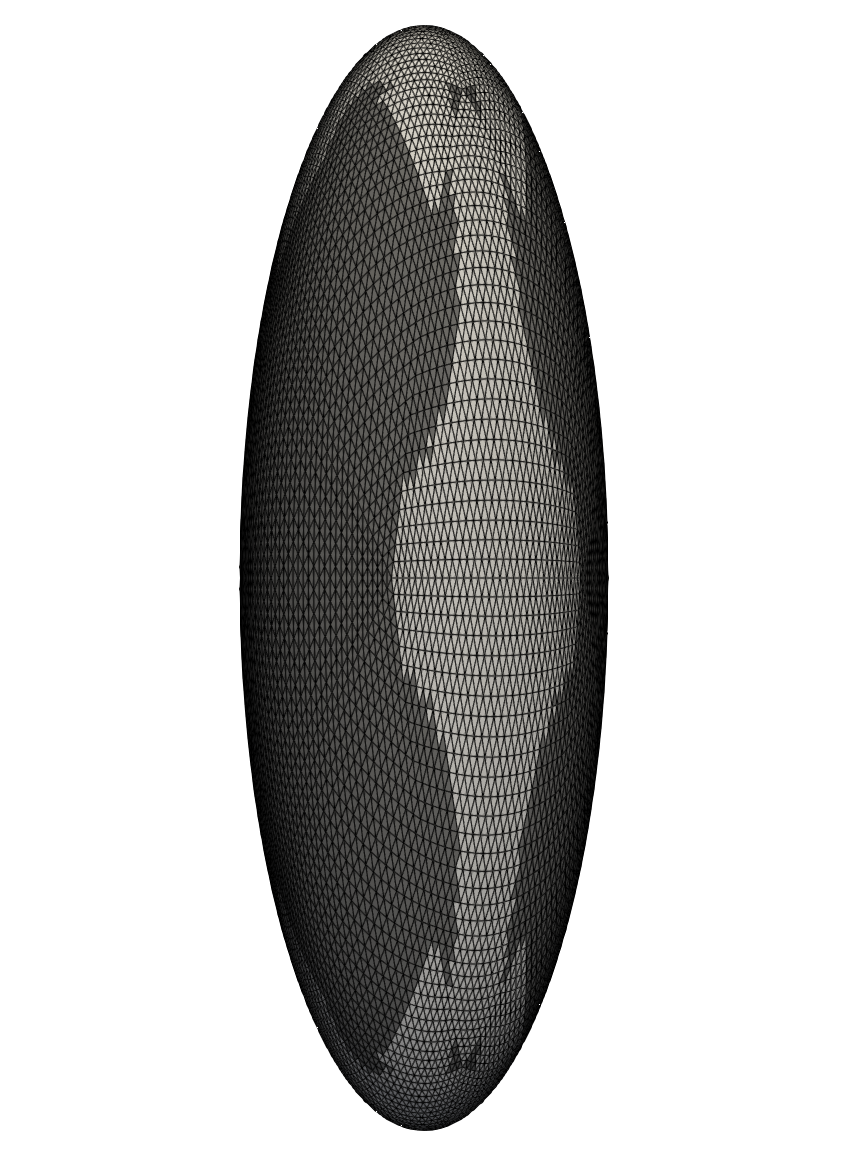}
	\caption{Discretized ellipsoid. Non-well centered triangles are darkened}
	\label{fig:ellipsoid_grid}
\end{figure}

The mesh is built from a well-centered icosahedral discretization of the unit shpere, containing 20480 triangles. This icosphere is then shrinked and streched to fit the shape of the ellipsoid. At the end, only 40 percent of the triangles remain well centered. The final grid is shown in figure \ref{fig:ellipsoid_grid}.

A barycentric dual is used. The dual cells are generally non-flat. For example, the geometric dual of an edge $e$ shared by two triangles $T_1$ and $T_2$ is the union of the segments $[c_{T_1},c_e]$ and $[c(e),c_{T_2}]$ where $c_e$, $c_{T_1}$ and $c_{T_2}$ are the respective barycenters. If $T_1$ and $T_2$ are non-coplanar, then these two segments are not parallel. Similarly, the dual of a vertex shared by non-coplanar triangles is an union of non-coplanar elementary triangles.

The analytical Hodge is used. %Note that its matrix is built as described in section \ref{sec:construction} on each primal triangle and then assembled, independently of the flatness of the mesh.
Note that the elementwise construction described in section \ref{sec:construction} does not need to be modified since each primal element (a triangle) is flat, independently of the flatness of the whole mesh.

\begin{figure}%[h]
	\centering
	\includegraphics[width=21mm]{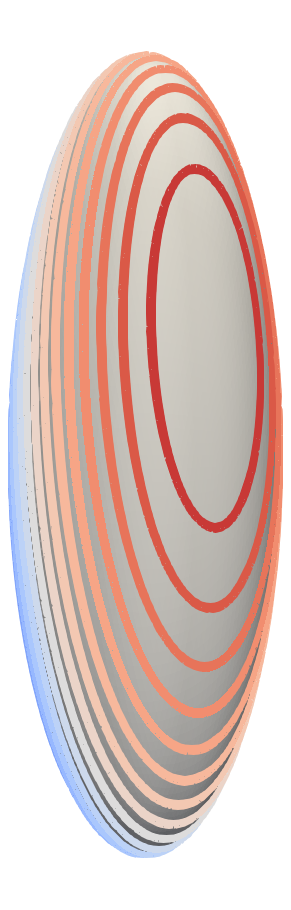}
	\includegraphics[width=21mm]{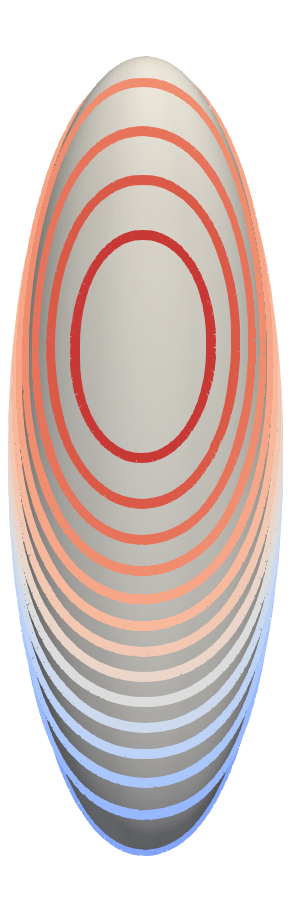}
	\includegraphics[width=21mm]{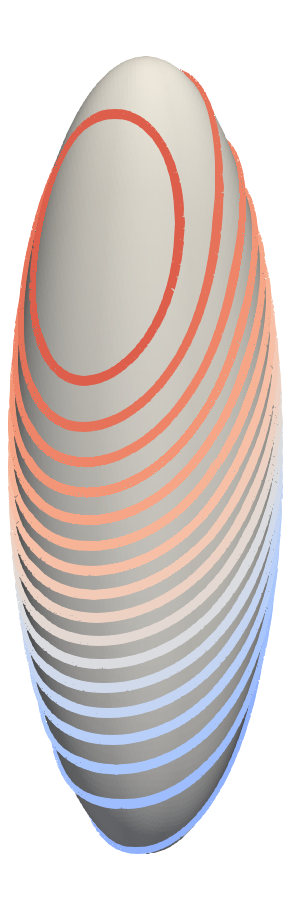}
	\includegraphics[width=21mm]{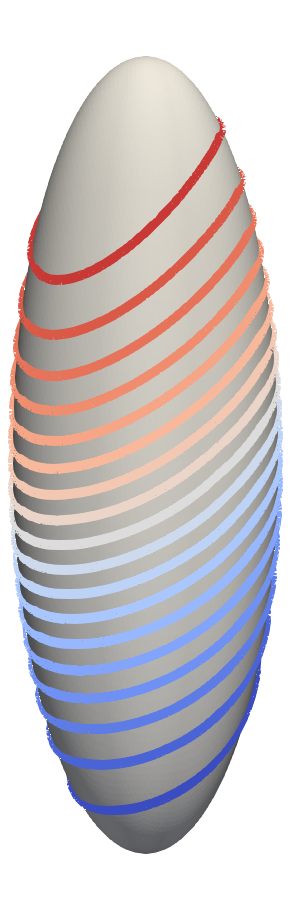}
	
	\includegraphics[width=21mm]{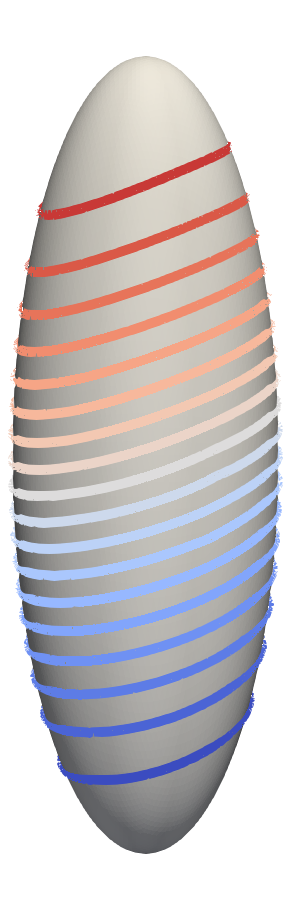}
	\includegraphics[width=21mm]{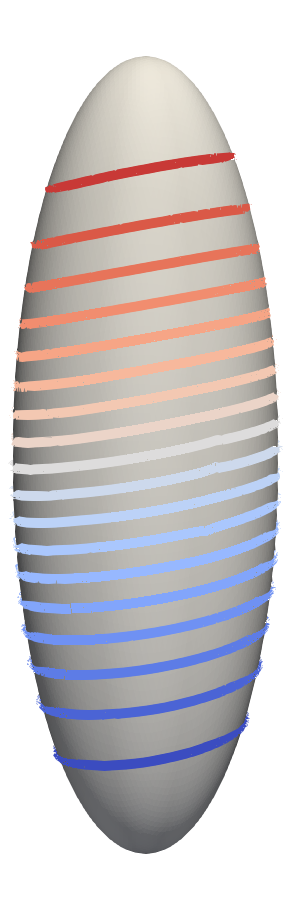}
	\includegraphics[width=21mm]{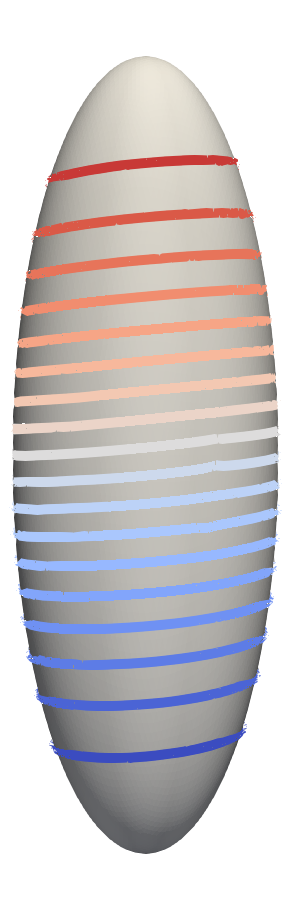}
	\includegraphics[width=21mm]{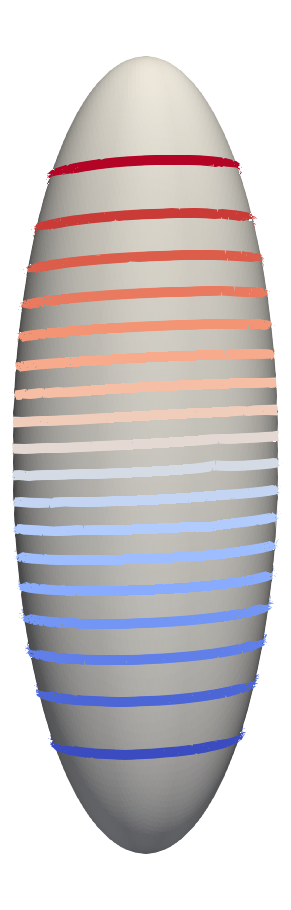}
	
	\includegraphics[width=21mm]{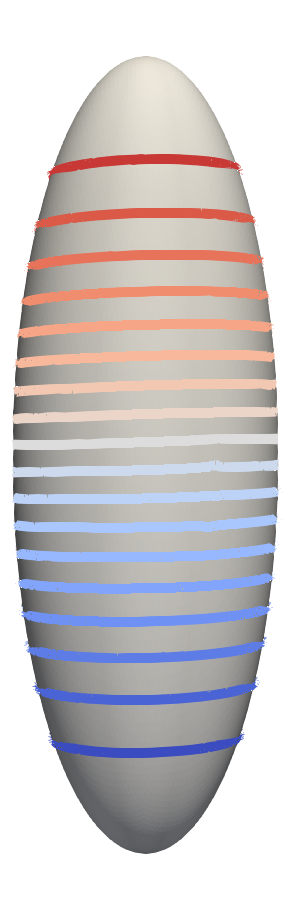}
	\includegraphics[width=21mm]{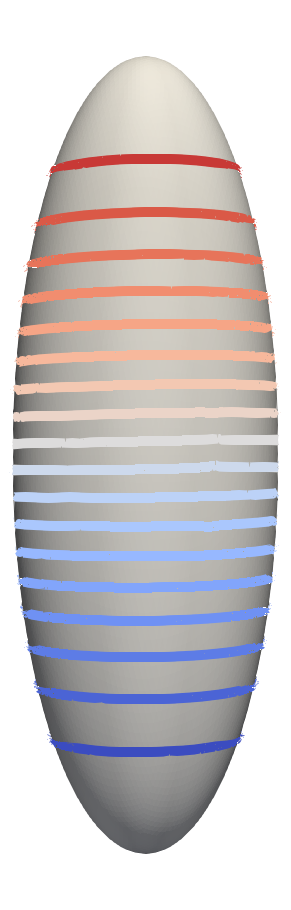}
	\includegraphics[width=21mm]{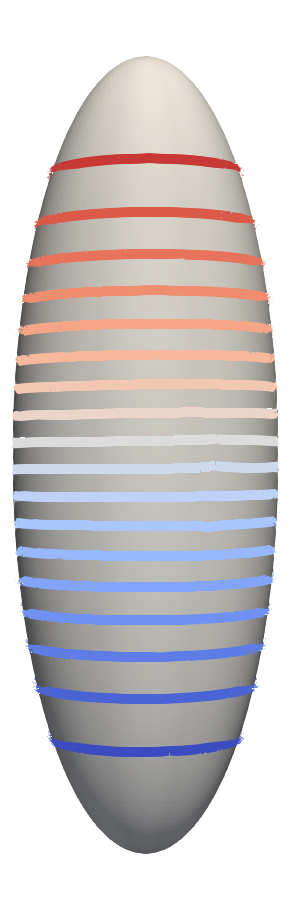}
	\includegraphics[width=21mm]{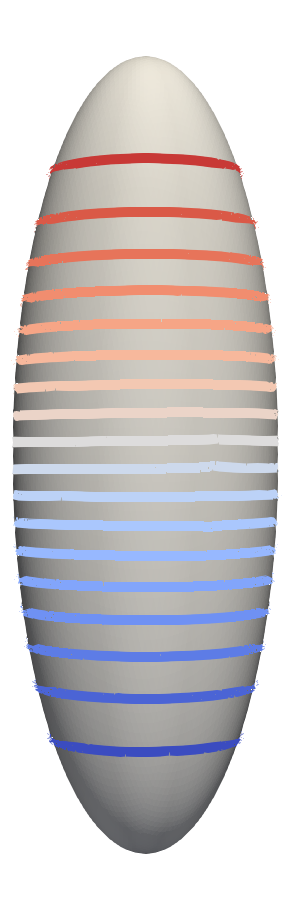}
	\caption{Streamlines for $t$ from 0 to 11}
	\label{fig:ellipsoid}
\end{figure}

Figure \ref{fig:ellipsoid} shows the streamlines at regular interval for $t$ from 0 to 11. As can be seen, the center of the vortex revolves around the ellipsoid untill it reaches a pole, where the curvature is the highest. The flow converges into a steady state. This result is in agreement with that obsereved in \cite{nitschke17}. 

This last test shows the potential of DEC and the new discrete Hodge in solving problems in non-flat domains.

\section{Conclusion}

We proposed a new construction of discrete Hodge operator in the context of DEC. It requires neither a well-centered mesh nor a circumcentric dual as with the diagonal Hodge. It also does not require a barycentric dual as in the geometric Hodge. The new Hodge is exact for piecewise constant forms, whichever interior points are chosen as centers. 

As shown by experiments with Poisson's equation, even in a well-centered mesh, the popular diagonal Hodge build on an orthogonal dual does not always give more accurate results than the new discrete Hodge. In the two simulations, the convergence rate of the new Hodge differs from that of the diagonal Hodge only by less than 0.51 percent.
Next, the new discrete Hodge competes well with the diagonal Hodge, in terms of convergence rate, in the resolution of the non-linear Navier-Stokes equations.  

The numerical tests in sections \ref{sec:anisothermal} and \ref{sec:nonstructured} showed the ability of DEC, and of the new discrete Hodge in particular, to solve problems involving thermal transfer.

Tests on various type of flat meshes were carried out. Structured acute (well-centered) meshes and right meshes were chosen in sections \ref{sec:poisson} to \ref{sec:anisothermal}. Unstructured but rather regular meshes, containing about 40 percent of non-Delaunay triangles were considered in section \ref{sec:nonstructured}. Sets of unstructured meshes with very irregular triangle shape, and comprising 15\%, 25\% or 50\% of non-Delaunay triangles were also used. In all of these configurations, the new Hodge star operator converges well. 

The test in section \ref{sec:surface} shows the ability of the discrete Hodge in handling non-flat domains. Note that the elemental construction proposed in section \ref{sec:construction} is independent of the flatness of the mesh.

Experiments in sections \ref{sec:poisson} to \ref{sec:anisothermal} tend to show that a barycentric dual provides as accurate as, or better results than an incentric dual. However, this should not be considered as the general case. Indeed, as shown in section \ref{sec:examples}, the incentric dual may be much more precise than the barycentric one in some cases. Moreover, in other types of problem, there may be other criteria than the error to take into consideration. 

Barycenter and incenter were chosen as centers of the dual meshes because of their simplicity. With our general construction, other interior points may also be considered, without breaking the exactness of the discrete Hodge on piecewise constant differential forms. The choice of the interior points may be done algorithmically to minimize some error indicator (for example, the overall error of the discerete Hodge operator on some class of differential forms). The development of such optimization process is curently under study by the authors.

Lastly, extension of the new discrete Hodge operator to the three-dimensional case is under construction.

%\clearpage
%\bibliographystyle{AIMS}
%\bibliography{../biblio}

\begin{thebibliography}{10}

\bibitem{Alotto04}
\newblock P.~Alotto and I.~Perugia,
\newblock Matrix properties of a vector potential cell method for
  magnetostatics,
\newblock \emph{IEEE Transactions on Magnetics}, \textbf{40} (2004),
  1045--1048.

\bibitem{arnold18}
\newblock D.~Arnold,
\newblock \emph{Finite Element Exterior Calculus},
\newblock SIAM-Society for Industrial and Applied Mathematics, 2018.

\bibitem{arnold06}
\newblock D.~Arnold, R.~Falk and R.~Winther,
\newblock Finite element exterior calculus, homological techniques, and
  applications,
\newblock \emph{Acta Numerica}, \textbf{15} (2006), 1–155.

\bibitem{arnold10}
\newblock D.~Arnold, R.~Falk and R.~Winther,
\newblock Finite element exterior calculus: from {H}odge theory to numerical
  stability,
\newblock \emph{Bulletin of the American Mathematical Society}, \textbf{47}
  (2010), 281--354.

\bibitem{Auchmann06}
\newblock B.~Auchmann and S.~Kurz,
\newblock A geometrically defined discrete {H}odge operator on simplicial
  cells,
\newblock \emph{IEEE Transactions on Magnetics}, \textbf{42} (2006), 643--646.

\bibitem{blanes16}
\newblock C.~Blanes~S.;~Fernando,
\newblock \emph{A Concise Introduction to Geometric Numerical Integration},
\newblock Monographs and Research Notes in Mathematics, CRC Press, 2016.

\bibitem{bochev06}
\newblock P.~Bochev and J.~Hyman,
\newblock \emph{Compatible Spatial Discretizations}, vol. 142 of The IMA
  Volumes in Mathematics and its Applications, chapter Principles of Mimetic
  Discretizations of Differential Operators, 89--119,
\newblock Springer, 2006.

\bibitem{bossavit88}
\newblock A.~Bossavit,
\newblock Whitney forms: A class of finite elements for three-dimensional
  computations in electromagnetism,
\newblock \emph{Science, Measurement and Technology, IEE Proceedings A},
  \textbf{135} (1988), 493--500.

\bibitem{bossavit98}
\newblock A.~Bossavit,
\newblock \emph{Computational electromagnetism. {V}ariational formulations,
  complementarity, edge elements},
\newblock Academic Press, 1998.

\bibitem{bossavit98_1}
\newblock A.~Bossavit,
\newblock On the geometry of electromagnetism: (1) {A}ffine space,
\newblock \emph{Journal of the Japan Society of Applied Electromagnetics and
  Mechanics}, \textbf{6} (1998), 17--28.

\bibitem{bossavit98_2}
\newblock A.~Bossavit,
\newblock On the geometry of electromagnetism: (2) {G}eometrical objects,
\newblock \emph{Journal of the Japan Society of Applied Electromagnetics and
  Mechanics}, \textbf{6} (1998), 114--123.

\bibitem{bossavit98_3}
\newblock A.~Bossavit,
\newblock On the geometry of electromagnetism: (3) {I}ntegration, {S}tokes,
  {F}araday’s law,
\newblock \emph{Journal of the Japan Society of Applied Electromagnetics and
  Mechanics}, \textbf{6} (1998), 233--240.

\bibitem{bossavit98_4}
\newblock A.~Bossavit,
\newblock On the geometry of electromagnetism: (4) {M}axwell's house,
\newblock \emph{Journal of the Japan Society of Applied Electromagnetics and
  Mechanics}, \textbf{6} (1998), 318--326.

\bibitem{bossavit99_3}
\newblock A.~Bossavit,
\newblock Computational electromagnetism and geometry : (3): {C}onvergence,
\newblock \emph{Journal of the Japan Society of Applied Electromagnetics and
  Mechanics}, \textbf{7} (1999), 401--408.

\bibitem{bossavit99_1}
\newblock A.~Bossavit,
\newblock Computational electromagnetism and geometry: (1) {N}etwork equations,
\newblock \emph{Journal of the Japan Society of Applied Electromagnetics and
  Mechanics}, \textbf{7} (1999), 150--159.

\bibitem{bossavit99_2}
\newblock A.~Bossavit,
\newblock Computational electromagnetism and geometry: (2) {N}etwork
  constitutive laws,
\newblock \emph{Journal of the Japan Society of Applied Electromagnetics and
  Mechanics}, \textbf{7} (1999), 204--301.

\bibitem{bossavit99_4}
\newblock A.~Bossavit,
\newblock Computational electromagnetism and geometry : (4): {F}rom degrees of
  freedom to fields,
\newblock \emph{Journal of the Japan Society of Applied Electromagnetics and
  Mechanics}, \textbf{8} (2000), 102--109.

\bibitem{bossavit99_5}
\newblock A.~Bossavit,
\newblock Computational electromagnetism and geometry : (5): {T}he {G}alerkin
  {H}odge,
\newblock \emph{Journal of the Japan Society of Applied Electromagnetics and
  Mechanics}, \textbf{8} (2000), 203--209.

\bibitem{bossavit05}
\newblock A.~Bossavit,
\newblock Discretization of electromagnetic problems: The “generalized finite
  differences” approach,
\newblock in \emph{Numerical Methods in Electromagnetics}, vol.~13 of Handbook
  of Numerical Analysis,
\newblock Elsevier, 2005,
\newblock 105 -- 197.

\bibitem{cassidi81}
\newblock C.~Cassidy and G.~Lord,
\newblock A square acutely triangulated,
\newblock \emph{Journal of Recreational Mathematics}, \textbf{13}.

\bibitem{cinalli04}
\newblock M.~Cinalli, F.~Edelvik, R.~Schuhmann and T.~Weiland,
\newblock Consistent material operators for tetrahedral grids based on
  geometrical principles,
\newblock \emph{International Journal of Numerical Modelling: Electronic
  Networks, Devices and Fields}, \textbf{17} (2004), 487 -- 507.

\bibitem{crane13}
\newblock K.~Crane, F.~de~Goes, M.~Desbrun and P.~Schröder,
\newblock Digital geometry processing with discrete exterior calculus,
\newblock in \emph{ACM SIGGRAPH 2013 courses},
\newblock SIGGRAPH '13, ACM, 2013.

\bibitem{desbrun05}
\newblock M.~Desbrun, A.~Hirani, M.~Leok and J.~Marsden,
\newblock Discrete exterior calculus,
\newblock \emph{arXiv:math/0508341}.

\bibitem{elcott07}
\newblock S.~Elcott, Y.~Tong, E.~Kanso, P.~Schröder and M.~Desbrun,
\newblock Stable, circulation-preserving, simplicial fluids,
\newblock \emph{ACM Transactions on Graphics}, \textbf{26}.

\bibitem{fecko06}
\newblock M.~Fecko,
\newblock \emph{Differential geometry and Lie groups for physicists},
\newblock Cambridge University Press, 2006.

\bibitem{guezaine09}
\newblock C.~Geuzaine and J.-F. Remacle,
\newblock Gmsh: A 3-{D} finite element mesh generator with built-in pre- and
  post-processing facilities,
\newblock \emph{International Journal for Numerical Methods in Engineering},
  \textbf{79} (2009), 1309--1331.

\bibitem{gillette09}
\newblock A.~Gillette,
\newblock \emph{Notes on Discrete Exterior Calculus},
\newblock Technical report, University of Texas at Austin, 2009.

\bibitem{grinspun05}
\newblock E.~Grispun, P.~Schröder and M.~Desbrun,
\newblock Discrete differential geometry: An applied introduction,
\newblock in \emph{ACM SIGGRAPH 2005 course notes},
\newblock SIGGRAPH '05, ACM, 2005.

\bibitem{hairer06}
\newblock W.~Hairer, G.~Wanner and C.~Lubich,
\newblock \emph{Geometric Numerical Integration. {S}tructure-Preserving
  Algorithms for Ordinary Differential Equations},
\newblock 2nd edition,
\newblock Springer Series in Computational Mathematics, Springer, 2006.

\bibitem{hiptmair01}
\newblock R.~Hiptmair,
\newblock Discrete {H}odge operators,
\newblock \emph{Numerische Mathematik}, \textbf{90} (2001), 265--289.

\bibitem{hirani03}
\newblock A.~Hirani,
\newblock \emph{Discrete Exterior Calculus},
\newblock Phd thesis, California Institute of Technology, Pasadena, CA, USA,
  2003.

\bibitem{Hirani13}
\newblock A.~Hirani, K.~Kalyanaraman and E.~VanderZee,
\newblock Delaunay {H}odge star,
\newblock \emph{Computer-Aided Design}, \textbf{45} (2013), 540--544.

\bibitem{hirani15}
\newblock A.~Hirani, K.~Nakshatrala and J.~Chaudhry,
\newblock Numerical method for {D}arcy flow derived using discrete exterior
  calculus,
\newblock \emph{International Journal for Computational Methods in Engineering
  Science and Mechanics}, \textbf{16} (2015), 151--169.

\bibitem{lee12}
\newblock J.~Lee,
\newblock \emph{Introduction to Smooth Manifolds},
\newblock 2nd edition,
\newblock Graduate Texts in Mathematics, Springer, 2012.

\bibitem{marsden99}
\newblock J.~Marsden and T.~Ratiu,
\newblock \emph{Introduction to Mechanics and Symmetry. {A} Basic Exposition of
  Classical Mechanical Systems}, vol.~17 of Texts in Applied Mathematics,
\newblock 2nd edition,
\newblock Springer Verlag, 1999.

\bibitem{mohamed16b}
\newblock M.~Mohamed, A.~Hirani and R.~Samtaney,
\newblock Comparison of discrete {H}odge star operators for surfaces,
\newblock \emph{Computer-Aided Design}, \textbf{78} (2016), 118 -- 125.

\bibitem{mohamed16a}
\newblock M.~Mohamed, A.~Hirani and R.~Samtaney,
\newblock Discrete exterior calculus discretization of incompressible
  {N}avier–{S}tokes equations over surface simplicial meshes,
\newblock \emph{Journal of Computational Physics}, \textbf{312} (2016), 175 --
  191.

\bibitem{mohamed18}
\newblock M.~Mohamed, A.~Hirani and R.~Samtaney,
\newblock Numerical convergence of discrete exterior calculus on arbitrary
  surface meshes,
\newblock \emph{International Journal for Computational Methods in Engineering
  Science and Mechanics}, \textbf{19} (2018), 194--206.

\bibitem{morita01}
\newblock S.~Morita,
\newblock \emph{Geometry of Differential Forms},
\newblock American Mathematical Society, 2001.

\bibitem{Mullen11hot:hodge-optimized}
\newblock P.~Mullen, P.~Memari, F.~Goes and M.~Desbrun,
\newblock {HOT}: {H}odge-optimized triangulations,
\newblock \emph{ACM Trans. Graph. (SIGGRAPH)}.

\bibitem{nitschke17}
\newblock I.~Nitschke, S.~Reuther and A.~Voigt,
\newblock \emph{Discrete {E}xterior {C}alculus ({DEC}) for the Surface
  {N}avier-{S}tokes Equation},
\newblock Springer International Publishing, 2017.

\bibitem{abraham88}
\newblock A.~R, M.~JE and R.~R.,
\newblock \emph{Manifolds, Tensor Analysis, and Applications},
\newblock Springer-Verlag, 1988.

\bibitem{Rajan94}
\newblock V.~Rajan,
\newblock Optimality of the {D}elaunay triangulation in {$\mathbb{R}^d$},
\newblock \emph{Discrete \& Computational Geometry}, \textbf{12} (1994),
  189--202.

\bibitem{amses18}
\newblock D.~Razafindralandy, A.~Hamdouni and M.~Chhay,
\newblock A review of some geometric integrators,
\newblock \emph{Advanced Modeling and Simulation in Engineering Sciences},
  \textbf{5} (2018), 16.

\bibitem{amses19}
\newblock D.~Razafindralandy, V.~Salnikov, A.~Hamdouni and A.~Deeb,
\newblock Some robust integrators for large time dynamics,
\newblock \emph{Advanced Modeling and Simulation in Engineering Sciences},
  \textbf{6} (2019), 5.

\bibitem{salnikov18}
\newblock V.~Salnikov and A.~Hamdouni,
\newblock From modelling of systems with constraints to generalized geometry
  and back to numerics,
\newblock \emph{ZAMM - Journal of Applied Mathematics and Mechanics /
  Zeitschrift f\"ur Angewandte Mathematik und Mechanik}, \textbf{99} (2019),
  e201800218.

\bibitem{spivak991}
\newblock M.~Spivak,
\newblock \emph{A Comprehensive Introduction to Differential Geometry}, vol.~1,
\newblock 3rd edition,
\newblock Publish or Perish, 1999.

\bibitem{spivak992}
\newblock M.~Spivak,
\newblock \emph{A Comprehensive Introduction to Differential Geometry}, vol.~2,
\newblock 3rd edition,
\newblock Publish or Perish, 1999.

\bibitem{spivak993}
\newblock M.~Spivak,
\newblock \emph{A Comprehensive Introduction to Differential Geometry}, vol.~3,
\newblock 3rd edition,
\newblock Publish or Perish, 1999.

\bibitem{spivak994}
\newblock M.~Spivak,
\newblock \emph{A Comprehensive Introduction to Differential Geometry}, vol.~4,
\newblock 3rd edition,
\newblock Publish or Perish, 1999.

\bibitem{spivak995}
\newblock M.~Spivak,
\newblock \emph{A Comprehensive Introduction to Differential Geometry}, vol.~5,
\newblock 3rd edition,
\newblock Publish or Perish, 1999.

\bibitem{tarhasaari99}
\newblock T.~Tarhasaari, L.~Kettunen and A.~Bossavit,
\newblock Some realizations of a discrete {H}odge operator: a reinterpretation
  of finite element techniques,
\newblock \emph{IEEE Transactions on Magnetics}, \textbf{35} (1999),
  1494--1497.

\bibitem{vanderzee10}
\newblock E.~VanderZee, A.~Hirani, D.~Guoy and E.~Ramos,
\newblock Well-centered triangulation,
\newblock \emph{SIAM Journal on Scientific Computing}, \textbf{31}.

\bibitem{whitney57}
\newblock H.~Whitney,
\newblock \emph{Geometric integration theory},
\newblock Princeton University Press, 1957.

\bibitem{yuan10}
\newblock L.~Yuan,
\newblock Acute triangulations of trapezoids,
\newblock \emph{Discrete Applied Mathematics}, \textbf{158} (2010), 1121 --
  1125.

\bibitem{zamfirescu13}
\newblock C.~T. Zamfirescu,
\newblock Survey of two-dimensional acute triangulations,
\newblock \emph{Discrete Mathematics}, \textbf{313} (2013), 35 -- 49,
\newblock
  \urlprefix\url{http://www.sciencedirect.com/science/article/pii/S0012365X12004219}.

\end{thebibliography}

\providecommand{\href}[2]{#2}
\providecommand{\arxiv}[1]{\href{http://arxiv.org/abs/#1}{arXiv:#1}}
\providecommand{\url}[1]{\texttt{#1}}
\providecommand{\urlprefix}{URL }

\end{document}